\documentclass[fleqn,usenatbib]{mnras}

\usepackage{newtxtext,newtxmath}

\usepackage[T1]{fontenc}

\DeclareRobustCommand{\VAN}[3]{#2}
\let\VANthebibliography\thebibliography
\def\thebibliography{\DeclareRobustCommand{\VAN}[3]{##3}\VANthebibliography}

\usepackage[pdftex]{graphicx}
\usepackage{enumitem}
\usepackage{mathtools}
\usepackage[version=4]{mhchem}
\usepackage{siunitx}
\usepackage{booktabs}
\usepackage{amsmath,bm}
\usepackage{graphicx}

\DeclarePairedDelimiterX\braket[2]{\langle}{\rangle}{#1 \delimsize\vert #2}
\newcommand*\diff{\mathop{}\!\mathrm{d}}
\usepackage{float}
\usepackage{array,multirow,graphicx}
\usepackage[dvipsnames]{xcolor}
\usepackage{euscript}
\definecolor{mygray}{gray}{0.5}

\title[Global 21-cm: Foreground Polarisation]{Bayesian data analysis for sky-averaged 21-cm experiments with contamination from linearly polarised foregrounds}

\author[E. Shen et al.]{
Emma Shen,$^{1,2}$\thanks{E-mail: yhs24@cam.ac.uk}
Dominic Anstey,$^{1,2}$
Marta Spinelli, $^{4,5}$
Eloy de Lera Acedo$^{1,2}$
and Anastasia Fialkov$^{2,3}$
\\
$^{1}$Cavendish Laboratory, University of Cambridge, Cambridge, CB3 0HE, United Kingdom\\
$^{2}$Kavli Institute for Cosmology, Madingley Road, Cambridge, CB3 0HA, United Kingdom\\
$^{3}$Institute of Astronomy, University of Cambridge, Madingley Road, Cambridge CB3 0HA, United Kingdom\\
$^{4}$Observatoire de la Côte d’Azur, Laboratoire Lagrange, Bd de l’Observatoire, CS 34229, 06304 Nice cedex 4, France\\
$^{5}$Department of Physics and Astronomy, University of the Western Cape, Bellville, Cape Town, 7535, South Africa
}

\date{Accepted XXX. Received YYY; in original form ZZZ}

\pubyear{2024}

\begin{document}
\label{firstpage}
\pagerange{\pageref{firstpage}--\pageref{lastpage}}
\maketitle

\begin{abstract}
The precise measurement of the sky-averaged HI absorption signal between 50 and 200 MHz is the primary goal of global 21-cm cosmology. This measurement has the potential to unravel the underlying physics of cosmic structure formation and evolution during the Cosmic Dawn. It is, however, hindered by various non-smooth, frequency-dependent effects, whose structures resemble those of the signal. One such effect is the leakage of polarised foregrounds into the measured intensity signal: polarised foreground emission undergoes Faraday rotation as it passes through the magnetic fields of the interstellar medium, imprinting a chromatic structure in the relevant frequency range which complicates the extraction of the cosmological HI absorption feature. We investigate the effect of polarised Galactic foregrounds on extracting the global 21-cm signal from simulated data using REACH's data analysis pipeline; the Radio Experiment for the Analysis of Cosmic Hydrogen (REACH) is an experiment designed to detect the sky-averaged 21-cm HI signal from the early Universe using physically informed models. Using the REACH pipeline, we successfully recover an injected global 21-cm signal with an amplitude of approximately 0.16 K, centred between 80 and 120 MHz, achieving a low root-mean-square error (less than 30\% of the injected signal strength) in all the tested cases. This includes scenarios with simulated polarised Galactic diffuse emissions and polarised point source emissions, provided the overall polarisation fraction is below $\sim 3\%$. The linear mixing of contamination, caused by the superposition of multiple patches with varying strengths of Faraday rotation, produces patterns that are more distinct from the global signal. This distinction makes global signal recovery easier compared to contamination resulting from a single, slow oscillation pattern.
\end{abstract}




\section{Introduction}
The HI 21-cm absorption line is considered one of the most promising probes to study Cosmic Dawn and the Epoch of Reionisation (EoR), periods in cosmic history that remain relatively unexplored. The 21-cm line arises from the hyperfine transition of neutral hydrogen. The level of emission or absorption is determined by the relative populations of the upper and lower hyperfine states, which, within the epoch of interest, is primarily a result of X-ray heating and the Wouthuysen-Field effect \citep{wouth,field}; the HI gas spin temperature and kinetic temperature are coupled via Lyman-$\alpha$ radiation emitted by the first light sources. The redshifted HI 21-cm line therefore has the potential to unravel the physics of cosmic structure formation and evolution during Cosmic Dawn and the Epoch of Reionisation \citep{hogan79, madau97, barkana01, furl06, pri08}.

Currently existing global experiments include REACH \citep{natreach}, EDGES \citep{bowman08}, SARAS \citep{patra13, singh17}, LEDA \citep{ledaprice} PRIZM \citep{philip19}, DAPPER \citep{burns21}, and MIST \citep{mist24}. \citet{edges} reported a sky-averaged 21-cm signal detection in the shape of a \textit{flattened Gaussian} centred at 78 MHz with an amplitude of 500 mK in 2018, which exceeds expectations from standard astrophysical models. However, re-examinations \citep{hillsnat, singh} on their methods have raised concerns regarding the non-physical parameters, the non-uniqueness of their solution, and potentially unaccounted-for systematic structures in the data. Moreover, \citet{hillsnat}, \citet{sims} and \citet{bevins} have shown that a damped sinusoidal systematic is strongly preferred in the EDGES data. SARAS3 also reported non-detection of EDGES' best-fit profile with 95.3\% confidence \citep{singh22}.

Measuring the global 21-cm signal is challenging because the expected signal is extremely weak compared to the radio foregrounds, which are predominantly made up of galactic synchrotron radiation and free-free radiation. These foregrounds are approximately 4 orders of magnitude brighter than the signal. The intrinsic foreground intensity varies with frequency and follows a well-approximated power law. The foregrounds are therefore considered to be spectrally smooth in comparison to the signal. However, the presence of various non-smooth chromatic distortions makes it very difficult to fit the foregrounds by smooth functions alone \citep{bern09}. 

Polarised Galactic synchrotron emission can result in such distortions; polarised Galactic emissions are Faraday rotated when they pass through the magnetised interstellar medium \citep{ryb86}. The rotation in the polarised angle of emission is proportional to the square of its wavelength ($\lambda^2$), and thus the amount of contamination leaked into the total intensity varies with frequency. The observed polarised Galactic foregrounds are predominantly linear because the majority of electrons we do observe spiral around magnetic fields perpendicular to the line-of-sight (LOS) due to relativistic beaming. On the contrary, In contrast, observed circular polarisation is diminished because photons with opposite rotation directions cancel out when integrated along the line of sight \citep{torsten17}. 

Observation on the linear Galactic polarisation at the relevant low frequencies is currently lacking. The following are the existing few observations: Murchison Widefield Array (MWA) carried out a 2400 degree$^2$ survey at 189 MHz \citep{bern13}. The S-band Polarisation All-Sky Survey (S-PASS) has observed the entire southern sky with the 64-meter Parkes radio telescope at 2.3 GHz \citep{spass}. LOFAR Two-meter Sky Survey (LOTSS) produced a Faraday depth cube mosaic covering 568 degree$^2$ in the frequency range 120–168 MHz to map diffuse polarised synchrotron emission \citep{eck19}. Listed as one of its science goals \citep{skalow}, SKA-Low hopefully will bring a remedy to the lack of polarisation data within the next decade. 

Contamination due to polarised foregrounds for sky-averaged experiments has been studied by \citet{spin2018, spin2019}; they created a model to simulate contamination by polarised foregrounds through simulating all-sky polarised Galactic diffuse emission at low frequencies based on the statistical properties shown in the patchy MWA 2400 degree$^2$ survey. Faraday rotation measure synthesis was adopted to generate a 3D representation of the linearly polarised sky. In this paper, we study the effect of polarised foreground in the extraction of the global 21-cm signal feature. The models to simulate a linearly polarised sky include the one used in \citet{spin2018} as well as other methods built to describe the otherwise complex polarised sky by tunable parameters. 

We specifically test the robustness of the REACH pipeline against polarised foreground contamination using its improved Bayesian data analysis approach. REACH is a global experiment built to measure the HI 21-cm line based on Bayesian analysis. It distinguishes itself from other global 21-cm experiments by its physically informed models of the signal, the foreground, and the instrument itself \citep{natreach}. The simulated antenna temperature data are generated using the REACH dipole beam with the frequency range 50-150 MHz. The data are then analysed by the time-separated REACH data analysis pipeline \citep{dom2}, meaning data in all time bins are jointly fitted. The fitting is realised by the advanced Bayesian nested sampling algorithm \textsc{PolyChord} \citep{hand5b}.

In section \ref{sec2}, we describe how leakage from a polarised sky can be modelled and show the three different approaches we take to simulate linearly polarised foregrounds. The simulated linearly polarised foregrounds are then analysed by the REACH pipeline, which is detailed in section \ref{sec3}; one can find an introduction to the Bayesian nested sampling technique and mechanisms of the REACH data analysis pipeline, including how the foregrounds are fitted using physically informed models and time-separated analysis. The results are presented in section \ref{sec4}. Section \ref{seccon} concludes the work. 

\section{Simulating Observation with Linearly Polarised Foregrounds}
\label{sec2}
Observation made by an antenna is unavoidably polarised. The observed brightness temperature contains the intrinsic intensity as well as polarisation leakage, and its direction depends on the type and the pointing of the antenna. The leakage due to polarised foregrounds can be described by Stokes parameters, which are directly measurable quantities. The total brightness temperature observed by an antenna would be the linear combination of the intrinsic foreground intensity (Stokes $I$) and the leakage. The leakage is the combination of Stokes $Q$, $U$, and $V$, and the composition depends on the type of the receiver and the alignment between the polarised emission and the receiver (appendix \ref{append1}). There is currently no all-sky observation on Stokes $Q$, $U$, and $V$ maps in the relevant frequencies and the emission leakage due to polarisation is not well understood. We aim to be as general as possible and simplify the simulation to have tunable parameters under control to test the pipeline. Therefore, we consider three different models to simulate polarisation leakage by generating Stokes parameter maps. In this section, one can find the three different polarised foreground models adopted in this paper: a model based on data given by the MWA 2400 degree$^2$ survey (section \ref{secrmst}), another by creating an all-sky Stokes map directly from the rotation measure catalogue of point sources (section \ref{secrmtable}), and the last one by randomly assigning Faraday depth values in different sky regions (section \ref{secfdr}).

\subsection{Simulating Global Signal Observation}
The simulated sky temperature, $T_\mathrm{sky}$, consists of two components: the linearly polarised foregrounds, $T'_\mathrm{f}$ ($T_\mathrm{f} + T_{Q}$ where $T_\mathrm{f}$ is the total foreground intensity and, $T_{Q}$, is the linearly polarised component as defined in equation \ref{eqstxxtyy}), and the sky-averaged 21-cm signal, $T_\mathrm{21}$:
\begin{equation}
\begin{aligned}
T_\mathrm{sky} (\theta,\phi,\nu) = T'_\mathrm{f} + T_\mathrm{21}.
\end{aligned}
\label{eqfg2}
\end{equation}
The sky model, $T_\mathrm{f}$, at frequency, $\nu$, without simulated polarisation leakage, is generated by scaling an instance of the 2008 Global Sky Model \citep{gsm2008} at 230 MHz  by the equation
\begin{equation}
\begin{aligned}
T_\mathrm{f} (\theta,\phi,\nu) = \left( T_\mathrm{230} (\theta,\phi) - T_\mathrm{CMB} \right) \left(\frac{\nu}{230}\right)^{-\beta (\theta,\phi)} + T_\mathrm{CMB},
\end{aligned}
\label{eqfg1}
\end{equation}
with spectral index
\begin{equation}
\begin{aligned}
\beta (\theta,\phi) = \frac{\log\left( \frac{T_\mathrm{230}(\theta,\phi) - T_\mathrm{CMB}}{T_\mathrm{408}(\theta,\phi) - T_\mathrm{CMB}} \right)}{\log \left( \frac{230}{408}\right)},
\end{aligned}
\label{eqspec}
\end{equation}
describing a spectral index map  derived by mapping every pixel of a GSM instance at 408 MHz, $T_\mathrm{408}(\theta,\phi)$,  onto the corresponding pixel of a GSM instance at 230 MHz, $T_\mathrm{230}(\theta,\phi)$. $T_\mathrm{CMB}$ is the cosmic microwave background temperature, set at 2.725 K. A frequency of 230 MHz is chosen to avoid contamination by the redshifted global 21-cm signal. The simulated Stokes $Q$ map $T_Q$, detailed in section \ref{secsq}, is added to the plain foreground (equation \ref{eqstxxtyy}) to produce a foreground with polarisation leakage,  $T'_\mathrm{f}$:
\begin{equation}
\begin{aligned}
T'_\mathrm{f} (\theta,\phi,\nu) = T_\mathrm{f} (\theta,\phi,\nu) + T_Q(\theta,\phi,\nu).
\end{aligned}
\label{eqfgwp}
\end{equation}
Including the injected sky-averaged 21-cm signal, approximated as a Gaussian in this work, the simulated sky temperature equation (\ref{eqfg2}) is rewritten as: 
\begin{equation}
\begin{aligned}
T_\mathrm{sky} (\theta,\phi,\nu) = T_\mathrm{f} (\theta,\phi,\nu) + T_Q(\theta,\phi,\nu) + T_\mathrm{21}(\theta,\phi,\nu).
\end{aligned}
\label{eqfg33}
\end{equation}
The contaminated sky temperature is convolved with the beam, $D$, specifically the REACH dipole beam specifically in this paper. An additional component, $\hat{\sigma}$, is the uncorrelated Gaussian noise that accounts for the random antenna temperature noise. The complete equation for the simulated antenna temperature data for a single time bin is then
\begin{equation}
\begin{aligned}
T_\mathrm{data}(\nu) = \frac{1}{4\pi}\int_{0}^{4\pi} D(\theta,\phi,\nu) T_\mathrm{sky} (\theta,\phi,\nu) d\Omega + \hat{\sigma},
\end{aligned}
\label{eqfore}
\end{equation}
where $\hat{\sigma}$ is the uncorrelated Gaussian noise. The resulting $T_\mathrm{data}$ represents the simulated dipole antenna temperature, including polarisation leakage and the injected signal. In this paper, each dataset consists of 24 hours of data collected at 5-minute intervals over three nights (8 hours per night).

\subsection{Simulating Stokes Parameter Sky Map}
\label{secsq}
In this section, we detail the three adopted models for all-sky Stokes $Q$ map generation. The models we adopt are motivated for introducing complexities that are under control: they are used to create  extreme conditions by tuning quantified parameters to test the robustness of the adopted data analysis pipeline. The effect from a realistic polarised sky is expected to be much milder than most of the cases we present.

To generate foregrounds with leakage due to linear polarisation, the modelled all-sky Stokes $Q$ map is  added to the Stokes $I$ (total intensity) map. The dipole beam employed in REACH has a frequency range of 50 to 150 MHz with 1 MHz resolution. As such, all the simulated Stokes $Q$ sky maps in this paper adopt the same frequency range. 

\subsubsection{Stokes Q Map by Rotation Measure Synthesis Technique}
\label{secrmst}
The first model tested consists of the polarised Galactic diffuse emission simulation presented in \citep{spin2018}. This simulation is based on the Faraday rotation measure synthesis technique, a method used to recover the intrinsic polarisation properties of polarised sources along a line of sight, generating all-sky Stokes $Q$ maps at sky direction $\hat{n}$ and frequency $\nu$. This model uses the MWA 2400 degree$^2$ survey at 189 MHz as its foundation. The interferometric observation originally sampled up only to degree angular scales, but was extrapolated to tens of degrees scales in the simulation to make it applicable to global experiments.

The Faraday rotation measure synthesis technique employs the Fourier relationship between the observed polarisation properties and a function describing the intrinsic polarisation, the Faraday dispersion function, to separate the polarised sources along a line-of-sight.

The rotation yielded by Faraday rotation as polarised synchrotron emission travels through the magnetised interstellar medium is given by the equation
\begin{equation}
\begin{aligned}
\chi(\lambda^2) = \chi_0 + \psi\cdot\lambda^2,
\end{aligned}
\label{eqrm}
\end{equation}
where $\chi_0$ is the intrinsic polarisation angle, and $\chi(\lambda^2)$ is the observed polarisation angle at wavelength $\lambda$.  Faraday depth is represented by $\psi$. The Faraday depth of a source is calculated by integrating along a line-of-sight towards the source \citep{burn}:
\begin{equation}
\begin{aligned}
\psi \propto \int^{\mathrm{source}}_0 \diff r n_e B_{\parallel},
\end{aligned}
\label{eqfdepth}
\end{equation}
where $n_e$ is the free electron density, $B_{\parallel}$ is the magnetic field component along the line of sight, and $dr$ represents the infinitesimal path length. Multiple sources with varying Faraday depths may exist along the same line of sight. In contrast, the rotation measure (RM) is commonly defined as the gradient of a polarisation angle, $\chi$:
\begin{equation}
\begin{aligned}
\mathrm{RM} = \frac{d\chi(\lambda^2)}{d\lambda^2}.
\end{aligned}
\label{eqrotm}
\end{equation}
The complex Faraday dispersion function, $\tilde{P}(\psi)$, is defined as 
\begin{equation}
\begin{aligned}
P(\lambda^2) = \int^{+\infty}_{-\infty} \tilde{P}(\psi)e^{2i\psi\lambda^2}\diff\psi,
\end{aligned}
\label{eqfrpair1}
\end{equation}
where $\tilde{P}(\psi)$ is the complex polarised surface brightness per unit Faraday depth, and $P(\lambda^2)$ is the complex surface brightness \citep{burn}. $P(\lambda^2)$ is physically meaningful only for $\lambda^2 \ge 0$. Its Fourier pair can then be written down:
\begin{equation}
\begin{aligned}
\tilde{P}(\psi) = \int^{+\infty}_{-\infty} {P}(\lambda^2)e^{-2i\psi\lambda^2}\diff\lambda^2.
\end{aligned}
\label{eqfrpair2}
\end{equation}
Via equation (\ref{eqfrpair1}), the simulated outputs at the frequencies of interest can be reconstructed from the Faraday rotation measure synthesis data that present the polarised intensity as a function of Faraday depth $\psi$. In this paper, we adopt the model that excludes the higher values of Faraday depth, $\psi > 5$  $\mathrm{rad}$ $ \mathrm{m}^{-2}$ (see the top panel in figure \ref{figregs}), as observations have shown that polarised emission in the low-frequency band is low at high Faraday depths \citep{haver04, bern09, lenc16}.

\subsubsection{Rotation Measure Map of Point Sources}
\label{secrmtable}
Presented by \citet{vaneck23}, RMTable2023 is a consolidated catalogue of rotation measures from point sources collected from 42 published catalogues. We use this catalogue to simulate another all-sky Stokes $Q$ map. The catalogue itself consists of data from discrete radio sources and does not cover the whole sky, so the rotation measures of the areas without data are interpolated using the nearest-neighbour interpolation. The middle panel in figure \ref{figregs} shows the interpolated all-sky map. To create the all-sky Stokes $Q$ map, the value of the Stokes $Q$ in each pixel is calculated by:
 \begin{equation}
\begin{aligned}
Q = I p \cos{(2\psi\lambda^2)},
\end{aligned}
\label{eqqip}
\end{equation}
where $I$ is the total intensity given by the scaled 2008 Global Sky Model \citep{gsm08}, and $p$ is the linear polarisation fraction. The formalism of the equation is motivated by the $\lambda^2$ dependence in the Faraday rotation angle of synchrotron emission (equation \ref{eqrm}). The total intensity of an emission includes both polarised and unpolarised components; $p$, the linear polarisation fraction, is the ratio of these components, and is constrained by $0 \leq p \leq 1$.

\subsubsection{Maps with Different Numbers of Faraday Depth Region}
\label{secfdr}
We present a toy model where the sky is divided into different numbers of regions with randomly assigned Faraday depths in each region to investigate how the complexity of  Faraday depth distribution would affect foreground polarisation and, in turn, signal extraction. The distribution of Faraday depths in a true sky is much more complicated than just a few regions. The purpose of this model is to study the effect of the Faraday depth distribution complexity on the  polarisation leakage. To introduce structures that are more difficult to distinguish from the global 21-cm signal, the assigned Faraday depths are limited to very low values ($< 0.5$ rad m$^{-2}$). This would make the oscillatory behaviour of Stokes $Q$ as a function of frequency relatively slow and resemble the global signal. A realistic sky, nevertheless, would be a result of all Faraday depths. We are doing this to test the robustness of the pipeline by introducing extreme scenarios.

\begin{figure*}
    \centering
    \minipage{0.49\textwidth} \includegraphics[trim={0 0 0cm 0cm},clip,width=\linewidth]{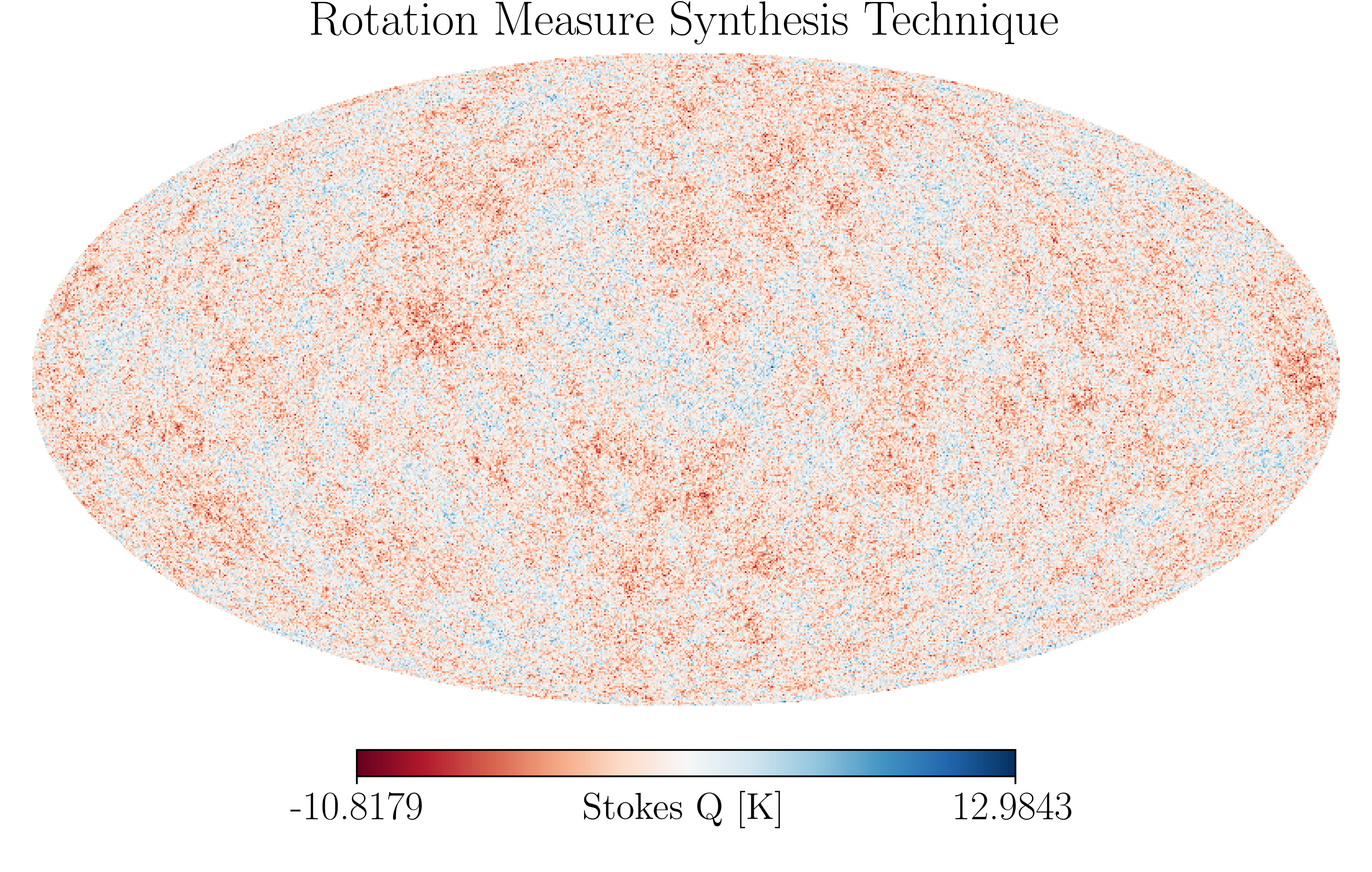}
    \endminipage\hfill \\[2.85mm]
    \centering
    \minipage{0.49\textwidth} \includegraphics[trim={0cm 0cm 0 0cm},clip,width=\linewidth]{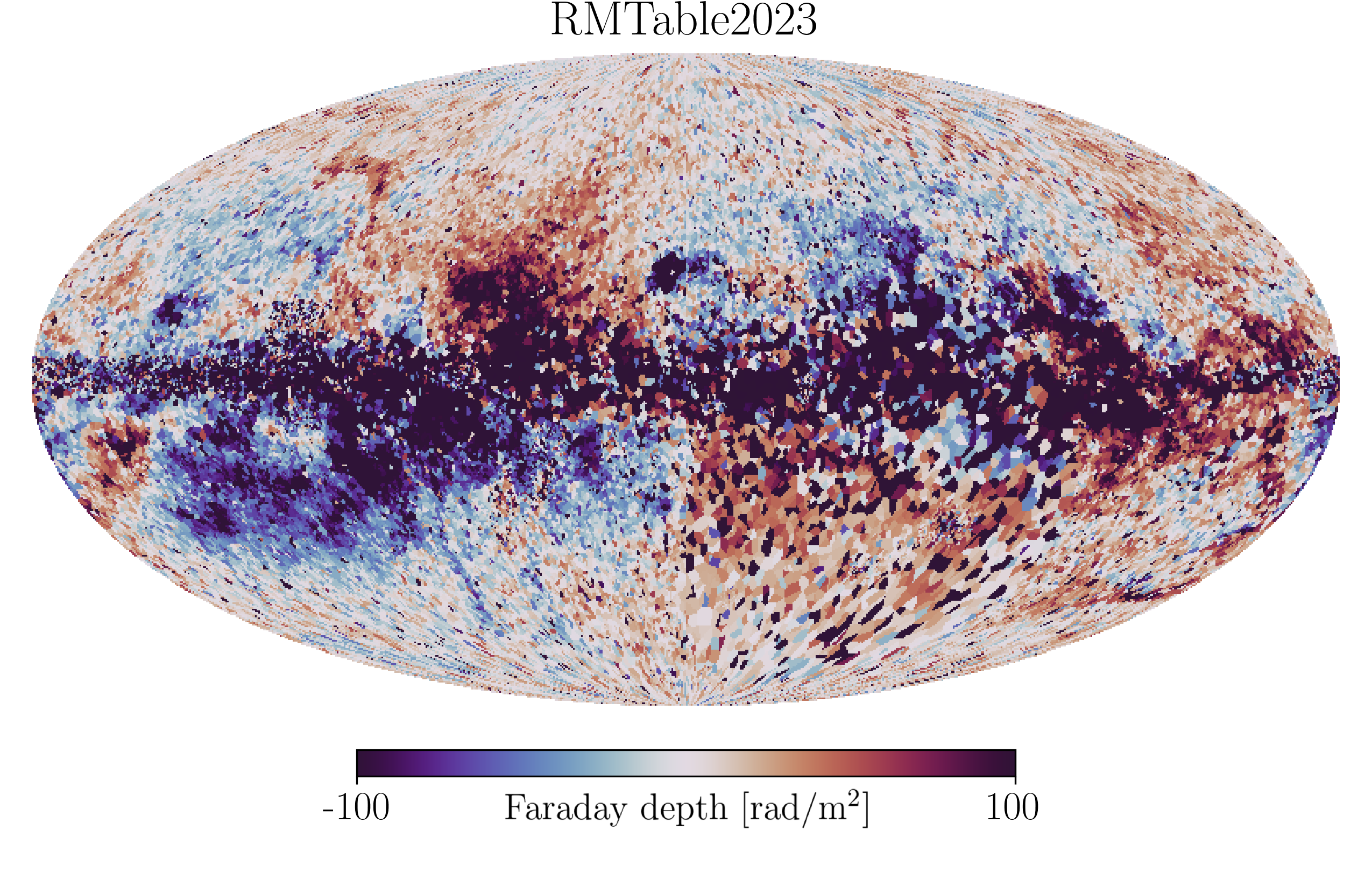}
    \endminipage\hfill\\[2.85mm]
    \centering
    \minipage{0.99\textwidth} \includegraphics[trim={0cm 0cm 0 0cm},clip,width=0.49\linewidth]{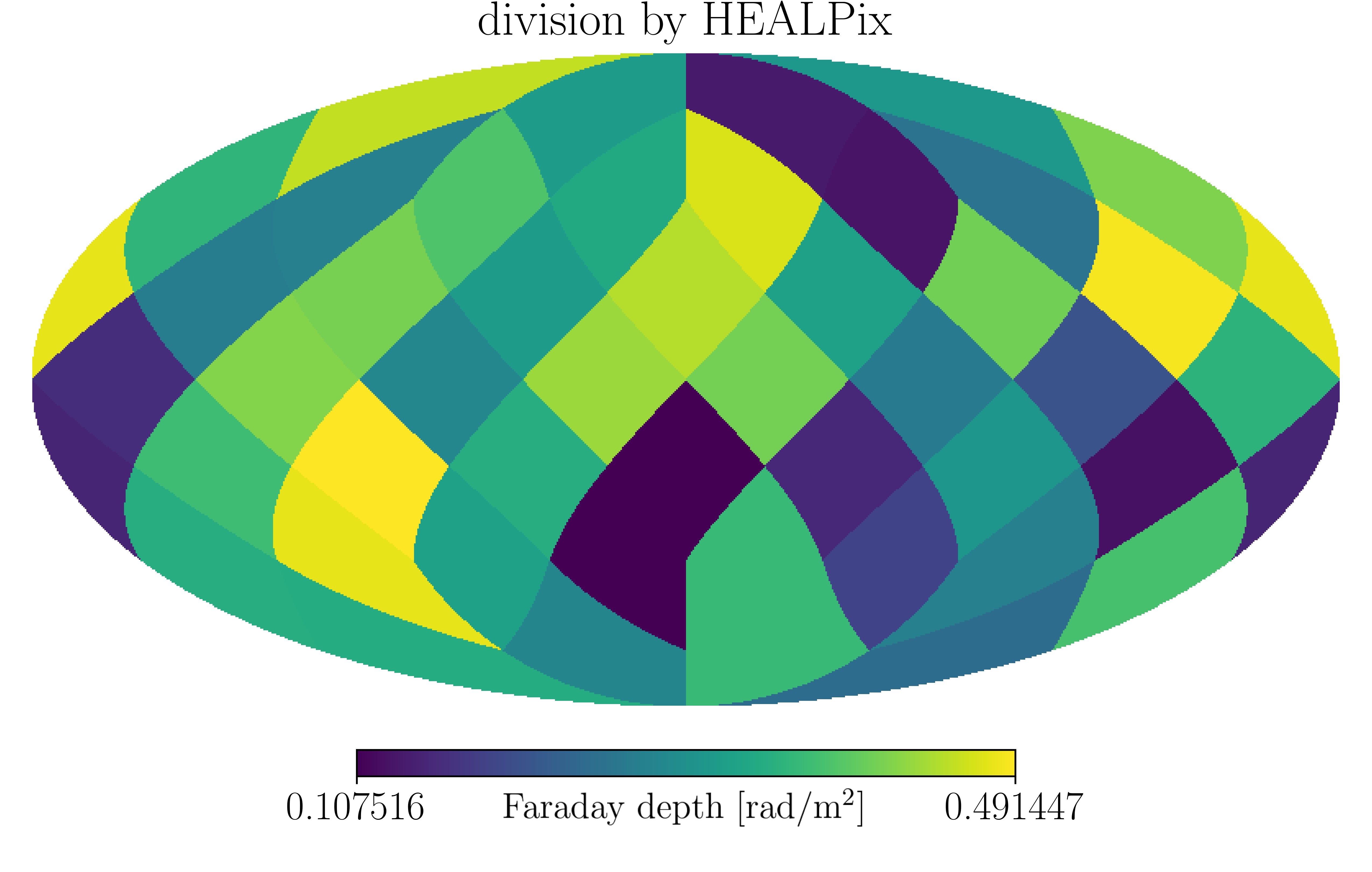}
    \includegraphics[trim={0cm 0cm 0 0cm},clip,width=0.49\linewidth]{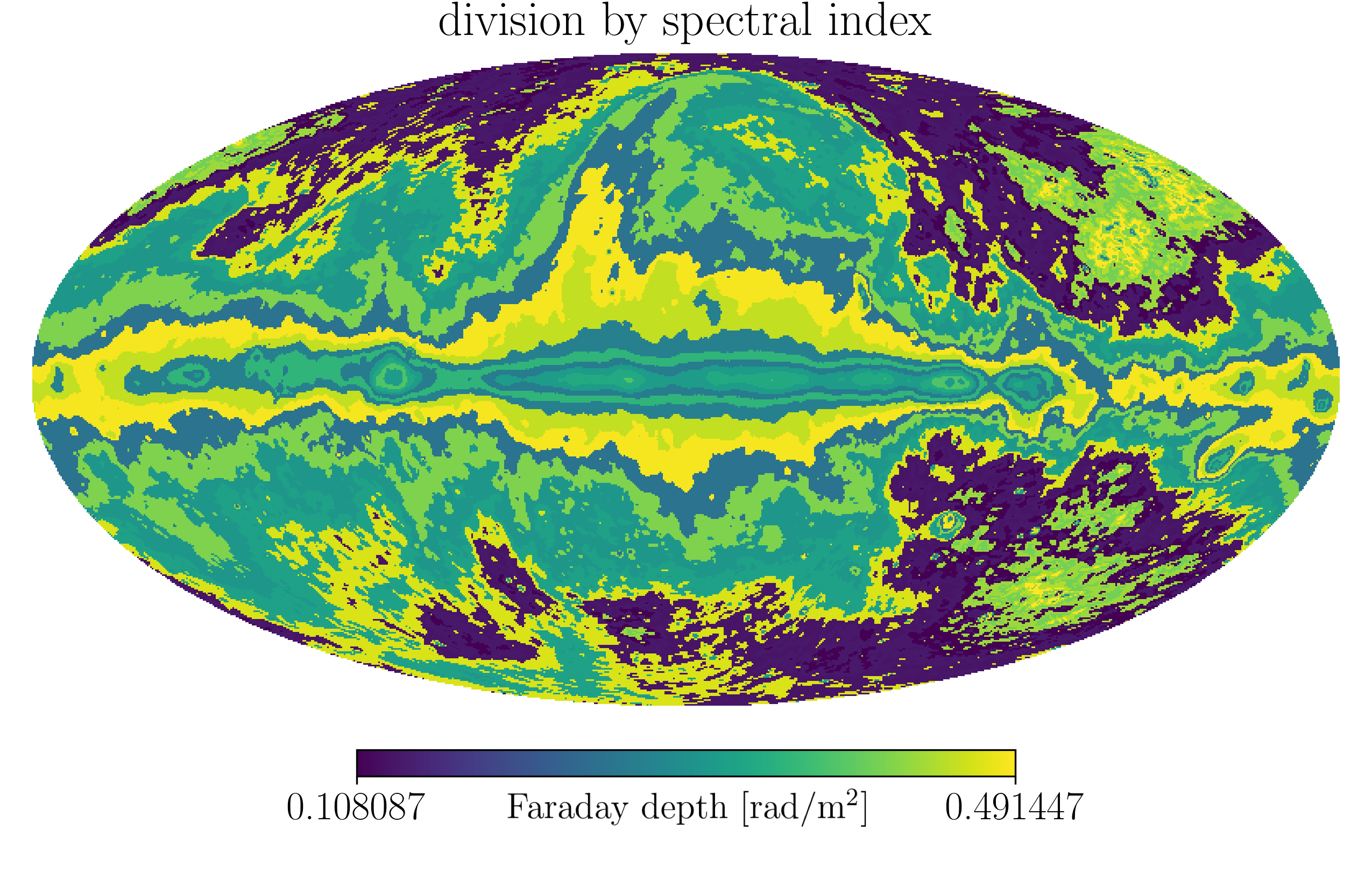}
    \endminipage\hfill
    \caption{The all-sky distribution of Stokes Q or Faraday depth value of the three different adopted models. The maps are in the Galactic coordinates. \textbf{Top: }the map on the top of the figure is the Stokes $Q$ map generated by the polarised Galactic diffuse emission simulation described in \citet{spin2018} and section \ref{secrmst}. In this model,  higher values of Faraday depth, $\psi > 5$  $\mathrm{rad}$ $ \mathrm{m}^{-2}$, are excluded as observations have shown that polarised emission in the low-frequency band has little emission at high Faraday depths \citep{haver04, bern09, lenc16}. It yields a relatively slower oscillation with respect to frequency compared to cases where all Faraday depths are taken into account. \textbf{Middle: } the middle one is the Faraday depth map generated based on the catalogue of Faraday rotation measure of radio sources provided by \citet{vaneck23}. The catalogue itself consists of data from discrete polarised radio sources and does not cover the whole sky, so the Faraday depth of the areas without data is interpolated using the nearest-neighbour interpolation. \textbf{Bottom: } the two maps at the bottom show the sky divided into regions with different values of Faraday depth ($<0.5$ rad m$^{-2}$) assigned randomly. \textbf{Bottom-left: } the one on the left is divided into 48 ($\mathrm{N_{side}}=2$) regions via HEALPix's gridding scheme. \textbf{Bottom-right: } the one on the right is divided into 25 regions using the REACH data analysis pipeline, which is based on the spectral indices.}
    \label{figregs}
\end{figure*}

The sky is divided in two different ways: the first one is simply using HEALPix's gridding scheme and dividing the whole sky into equally sized 12 ($\mathrm{N_{side}}=1$), 48 ($\mathrm{N_{side}}=2$), and 196 ($\mathrm{N_{side}}=4$) regions respectively. The selected numbers of sides are the lowest possible ones for HEALPix gridding. The second one takes the sky division scheme based on the spectral indices from the REACH data analysis pipeline. The size of each region varies depending on the local spectral index distribution \citep{dom1}. The bottom panels in figure \ref{figregs} show one example for each case. Like in the previous model, the value of the Stokes $Q$ in each pixel in the all-sky Stokes $Q$ map is calculated by equation (\ref{eqqip}).

\section{Data Analysis Pipeline}
\label{sec3}
This work employs the REACH data analysis pipeline to extract an injected global 21-cm signal from polarisation-contaminated antenna temperature data, generated as described in section \ref{sec2}. As detailed in \citet{dom1} and \citet{dom2}, the REACH data analysis pipeline is designed to reconstruct the redshifted global 21-cm signal using physically motivated foreground models \citep{natreach}, distinguishing it from standard pipelines. The adopted fitting algorithm is \textsc{PolyChord} \citep{hand5b}, an advanced Bayesian nested sampling algorithm, opted for its ability to effectively calculate Bayesian evidence for cases with high dimensionalities. The number of parameters in our analysis is large, making \textsc{PolyChord} essential to the REACH data analysis pipeline. In this section, we describe how we analyse the datasets using this Bayesian-inference-based pipeline.

\subsection{Bayesian Inference Based Fitting}
Bayesian inference is particularly valuable for statistical modelling due to its ability to perform both model comparison and parameter estimation. It is realised by invoking Bayes' theorem: 
\begin{equation}
\begin{aligned}
\mathrm{P}(\theta_\mathcal{M}|\mathcal{D},\mathcal{M}) &= \frac{\mathrm{P}(\mathcal{D}|\theta_\mathcal{M},\mathcal{M})\mathrm{P}(\theta_\mathcal{M}|\mathcal{M})}{\mathrm{P}(\mathcal{D}|\mathcal{M})}.
\end{aligned}
\label{2}
\end{equation}
This equation shows how a model $\mathcal{M}$ parametrised by $\theta_\mathcal{M}$ updates prior knowledge of the parameters parameters $\mathrm{P}(\theta_\mathcal{M}|\mathcal{M})$ with data $\mathcal{D}$ to compute the posterior probability of $\theta_\mathcal{M}$. This equation can also be written in a cleaner way:
\begin{equation}
\begin{aligned}
\mathcal{P} = \frac{\mathcal{L}\pi}{\mathcal{Z}},
\end{aligned}
\label{eqbayes1}
\end{equation}
where $\mathcal{P}$ is the posterior distribution, $\mathcal{L}$ is the likelihood, the probability of the data given a model and the set of parameters describing the model, $\mathcal{\pi}$ is the prior distribution of the parameters, and $\mathcal{Z}$ is the Bayesian evidence, or marginal likelihood, which gives the probability of observing the data $\mathcal{D}$ given the model $\mathcal{M}$ marginalised over the parameters \citep{Sivia2006}. Given the same data, different models can be compared by Bayesian evidence:
\begin{equation}
\begin{aligned}
\mathrm{P}(\mathcal{M}|\mathcal{D}) = \frac{\mathrm{P}(\mathcal{D}|\mathcal{M})\mathrm{P}(\mathcal{M})}{\mathrm{P}(\mathcal{D})} = \mathcal{Z} \frac{\mathrm{P}(\mathcal{M})}{\mathrm{P}(\mathcal{D})},
\end{aligned}
\label{eqmodel}
\end{equation}
where $\mathrm{P}(\mathcal{D})$ is a normalisation factor independent of the model. Two models can then be compared by taking the ratio between two evidences weighted by the respective prior probabilities of the model $\mathrm{P}(\mathcal{M})$; it can be referred to as the Bayes factor.

Due to the need to marginalise over the full parameter space, calculating the Bayesian evidence can be difficult. Nevertheless, \textsc{PolyChord} utilises nested sampling \citep{skilling06}, which enables efficient estimation of the Bayesian evidence. The algorithm begins by randomly sampling the parameter space based on the given prior and calculating the likelihood for each sample. In each iteration, the point of the lowest likelihood is discarded and replaced by a new one of a greater likelihood. By this process, the prior volume of a parameter would contract exponentially towards the peak of the posterior.

\subsection{Physically Motivated Foreground Modelling}
The foreground model incorporates physically motivated simulations of foregrounds and antenna patterns to account for chromatic distortions during model fitting. It is realised by dividing the sky into regions within which the spectral indices are similar; a foreground function is generated by scaling a base all-sky map to each frequency, each region having a uniform spectral index. The resulting map is convolved with the antenna. The spectral indices of these regions become the parameters to be estimated. The model can become more refined to match the true sky by increasing the number of regions, with the required number guided by Bayesian evidence. This foreground model is especially robust against systematics, which is a crucial feature to the experiment. More detailed explanations can be found in \citet{dom1} and \citet{dom2}.

\subsection{Time-separated Analysis}
Each simulated dataset analysed in this study spans 24 hours, with 5-minute intervals between time bins, over three nights (8 hours per night). We adopt a foreground modelling approach that fits multiple datasets across different observation times (time bins) simultaneously, using a single Bayesian analysis without time integration. This creates a distinct foreground model for each time bin. With the additional information about the foregrounds and chromaticity, time-separated analysis allows the overall foreground modelling to perform at a much higher degree of accuracy, particularly for highly chromatic beams like the dipole beam \citep{dom2}. For our analysis, a Gaussian likelihood is appropriate since the simulated data includes Gaussian noise. In a time-integrated analysis, where both the data and the foreground model for each time bin are integrated over time, the Gaussian likelihood takes the form
\begin{equation}
\begin{aligned}
\log& \mathcal{L} = \sum_{i} -\frac{1}{2} \log (2\pi \sigma_\mathrm{n}^2) \\ &- \frac{1}{2} \left( \frac{\frac{1}{N_\mathrm{t}}\sum_j [T_\mathrm{data}(\nu_i,t_j)] - (\frac{1}{N_\mathrm{t}}\sum_j [T_\mathrm{F}(\nu_i,t_j,\theta_\mathrm{F})]+T_\mathrm{S}(\nu_i,\theta_\mathrm{S}))}{\sigma_\mathrm{n}}\right)^2
\end{aligned}
\label{eqlhint}
\end{equation}
for time bins $t_j$, where $T_\mathrm{data}$ is the observation data, $T_\mathrm{F}$ and $T_\mathrm{S}$ refer to the foreground and signal models respectively, with parameters $\theta_\mathrm{F}$ and $\theta_\mathrm{S}$, and $\sigma_\mathrm{n}$ is an additional parameter for the Gaussian noise. For a time-separated analysis, the resulting posterior values of the parameters should be the same for any time bin, so the likelihood can be rewritten as
\begin{equation}
\begin{aligned}
\log \mathcal{L} = & \sum_{i}\sum_{j} -\frac{1}{2} \log (2\pi \sigma_\mathrm{n}^2) \\ &- \frac{1}{2} \left( \frac{T_\mathrm{data}(\nu_i,t_j) - (T_\mathrm{F}(\nu_i,t_j,\theta_\mathrm{F})+T_\mathrm{S}(\nu_i,\theta_\mathrm{S}))}{\sigma_\mathrm{n}}\right)^2,
\end{aligned}
\label{eqlhsep}
\end{equation}
which means data of each time bin would be fitted to its corresponding foreground model to inform one shared set of parameters. A more thorough explanation can be found in \citet{dom2}.

\section{Results}
\label{sec4}
In this work, the signal root-mean-square error (RMSE) is utilised as a metric to evaluate the goodness of fit. "The signal RMSE is defined as the RMSE between the injected signal $T_{\mathrm{21}}$ and the reconstructed signal $T_{\mathrm{S}}$  and it is calculated by the following equation:
\begin{equation}
\begin{aligned}
\mathrm{RMSE} = \sqrt{\sum_{i=1}^{n}\frac{(T_{\mathrm{21}}(\nu_i) - T_{\mathrm{S}}(\nu_i))^2}{n}},
\end{aligned}
\label{eqrmse}
\end{equation}
where $\nu_i$ is $i^{\mathrm{th}}$ frequency bin within the frequency range. A lower signal RMSE corresponds to a better fit. In this paper, a recovered signal with signal RMSE lower than 30\% of the injected signal strength is considered a good fit with low signal RMSE. The results are divided into three sections based on the three different models introduced in section \ref{sec2}.

\subsection{Rotation Measure Synthesis Technique} 
\citet{spin2019} showed that the oscillatory behaviour of the Stokes $Q$ parameter as a function of frequency reveals notable differences between cases in which all Faraday depths are taken into account (`all $\phi$') and those restricted to low Faraday depth values ($\psi \leq 5$ rad m$^{-2}$, `low $\phi$'). In the low-frequency band ($\leq 100$ MHz), oscillatory structures arising from polarisation leakage are more complex and exhibit greater contamination overall in the `all $\phi$' cases than in the `low $\phi$' cases. Conversely, in the `low $\phi$' cases, the oscillations become smoother in the high-frequency band ($> 100$ MHz) as well as slower and larger in magnitude.

\begin{figure}
    \centering
    \minipage{0.49\textwidth} \includegraphics[trim={0cm 0cm 0 0cm},clip,width=\linewidth]{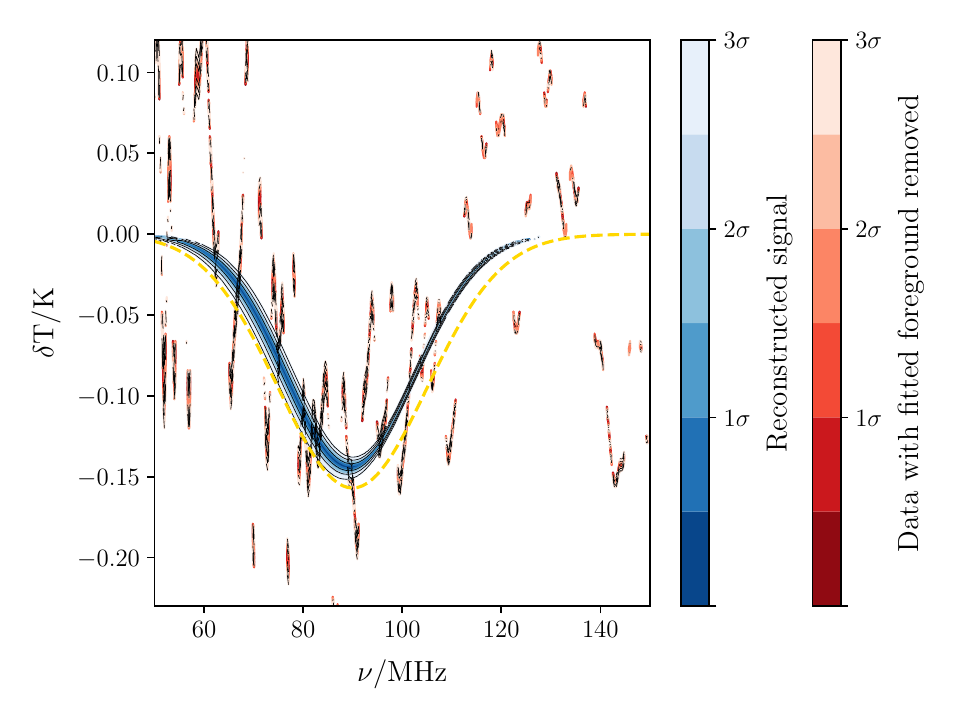}
    \endminipage\hfill
    \caption{The reconstructed signal (blue contours) and foreground residual (red contours) plots of an example case by the rotation measure synthesis technique. The yellow dashed line represents the injected signal with an amplitude of 0.157 K centred at 90 MHz. This example considers only low Faraday depth values ($\psi \leq 5$ rad m$^{-2}$). Visually, the reconstructed signal contours are already very close to the injected signal. The reconstructed signal RMSE is 0.0138 K, which is much lower than 30\% of the injected signal strength.}
    \label{figs75}
\end{figure}

We would like to know how the results change according to the position of the absorption feature with respect to the oscillatory pattern. For injected signals with an amplitude $A$ of 0.157 K and $\sigma =15$ MHz width centred at either 90 MHz or 120 MHz, no significant effects on the reconstructed signal are observed in cases where all Faraday depth values are included; however, foreground residuals do show significant difference in magnitude. In contrast, a standard pipeline would already be struggling with this level of contamination: in \citep{spin2018}, the amplitude of such contamination would have to be reduced for the standard pipeline to work. The same can be said for the `low $\phi$' cases with injected signal centred at 90 MHz; it is expected as the contamination in the low-frequency band is even lower. When the injected signal is centred at 120 MHz in the high-frequency band, a significant increase in residual magnitude can be observed. It is also expected, as the `low $\phi$' contamination in this frequency range oscillates at a rate that makes it resemble the injected signal, making signal recovery more challenging. The foreground temperature, by the power law, is also much lower at high frequencies, making the same contamination more prominent than at low frequencies. The time-separated pipeline is able to recover the signal with low signal RMSE (lower than 30\% of the injected signal strength) and high Bayes factor in all the tested cases. Figure \ref{figs75} shows one typical example of a `low $\phi$' case.
 
 \subsection{Rotation Measure Catalogue of Point Sources}
\begin{figure}
    \centering
    \minipage{0.49\textwidth} \includegraphics[trim={0cm 0cm 0 0cm},clip,width=\linewidth]{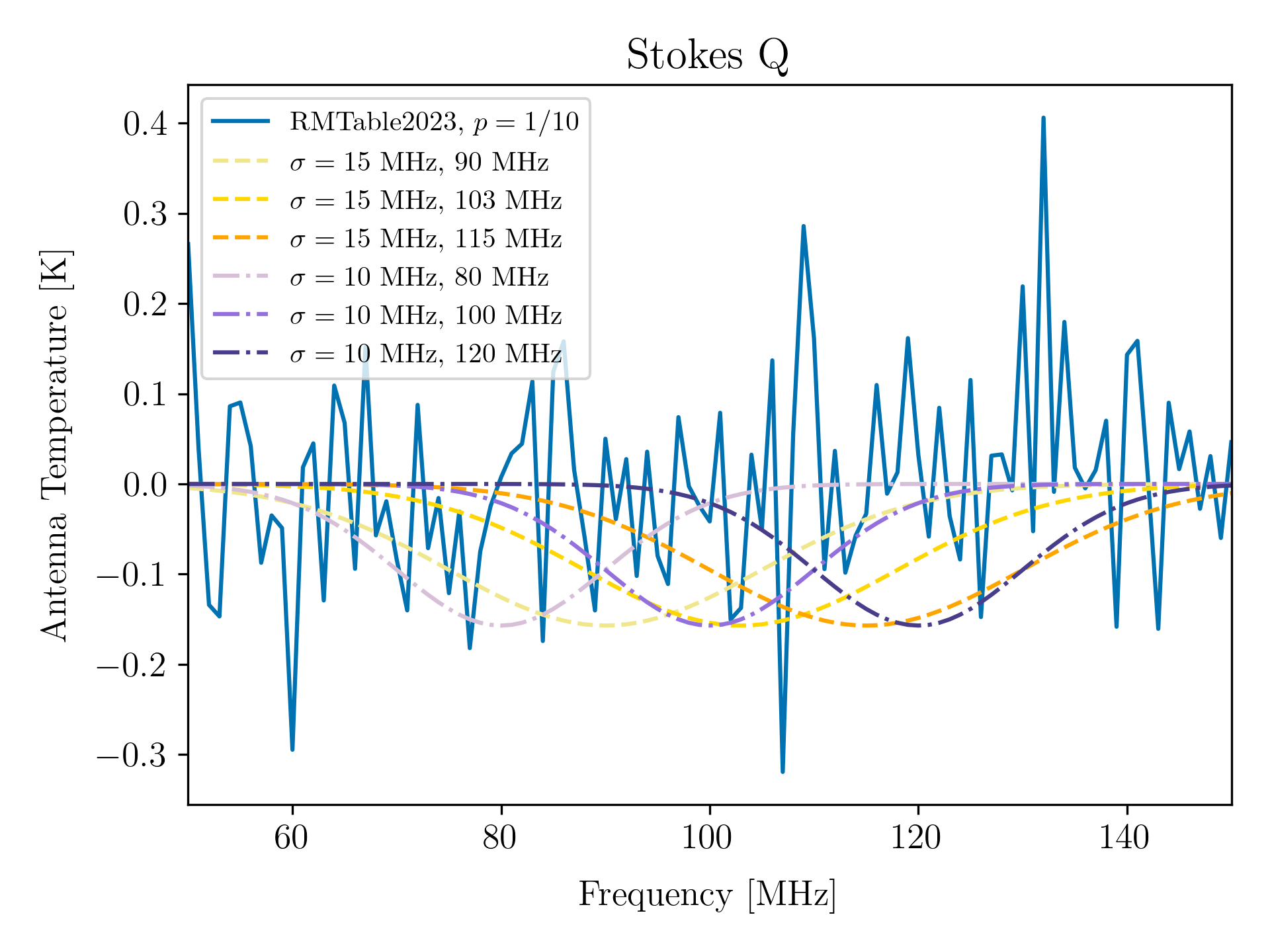}
    \endminipage\hfill
    \caption{Scaled Stokes $Q$ as a function of frequency. A uniform linear polarisation fraction of $p=1/10$ is applied to the whole sky at all frequencies. The Stokes $Q$ parameter map is simulated according to equation (\ref{eqqip}), and the Faraday depth is based on the RMTable2023 catalogue \citep{vaneck23}. The Stokes $Q$ structure in blue solid line is rather noisy compared to the injected global signals of 15 MHz width in yellow dashed lines and injected global signals of 10 MHz width in purple dashed lines.}
    \label{figrmt2023qmap}
\end{figure}

The all-sky Stokes $Q$ map used in this section is simulated based on the RMTable2023 catalogue given in \citet{vaneck23} via equation (\ref{eqqip}). The oscillation of the Stokes $Q$ parameter as a function of frequency, shown in figure \ref{figrmt2023qmap}, appears noisy. It can be explained by the distribution of many different Faraday depth values over the entire sky; the superposition of $\sin(a\lambda^2)$ waves oscillating at different rates will result in seemingly noisy structures. There are also a significant amount of regions with high Faraday depth values (middle panel in figure \ref{figregs}), which by equation (\ref{eqqip}) would intrinsically result in fast oscillation. For the dipole beam, the frequency range that allows for this test is not wide enough to test injected signals centred at too many different frequencies. The pipeline encounters difficulties in recovering the signal when the injected signal is centred near the edges of the frequency band; it often significantly overestimates the amplitude of the signal.

\begin{figure*}
    \centering
    \minipage{0.99\textwidth} \includegraphics[trim={0 0 5cm 0cm},clip,height=6.2cm]{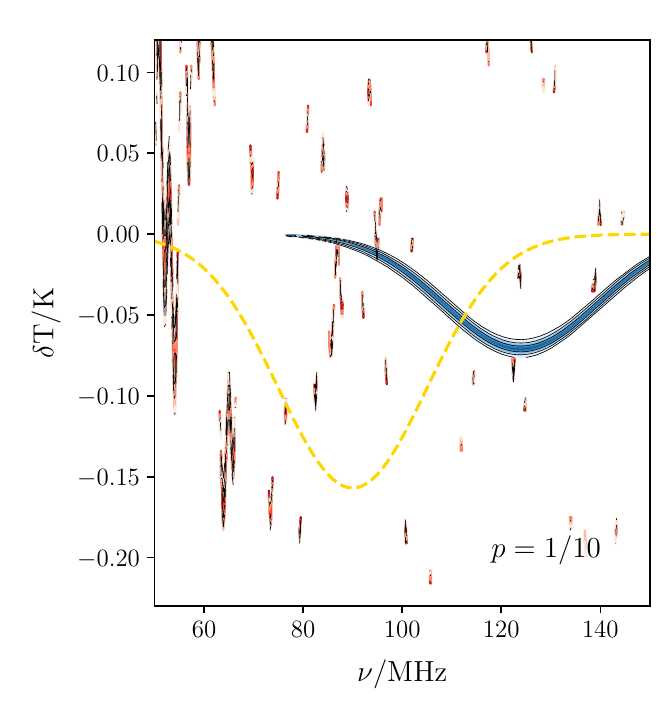}
    \includegraphics[trim={1.3cm 0 5cm 0cm},clip,height=6.2cm]{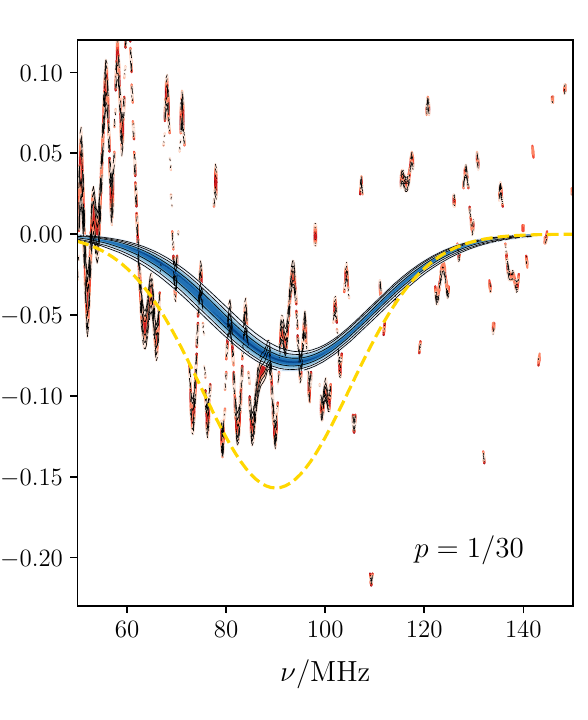}
    \includegraphics[trim={1.3cm 0 0.7cm 0cm},clip,height=6.2cm]{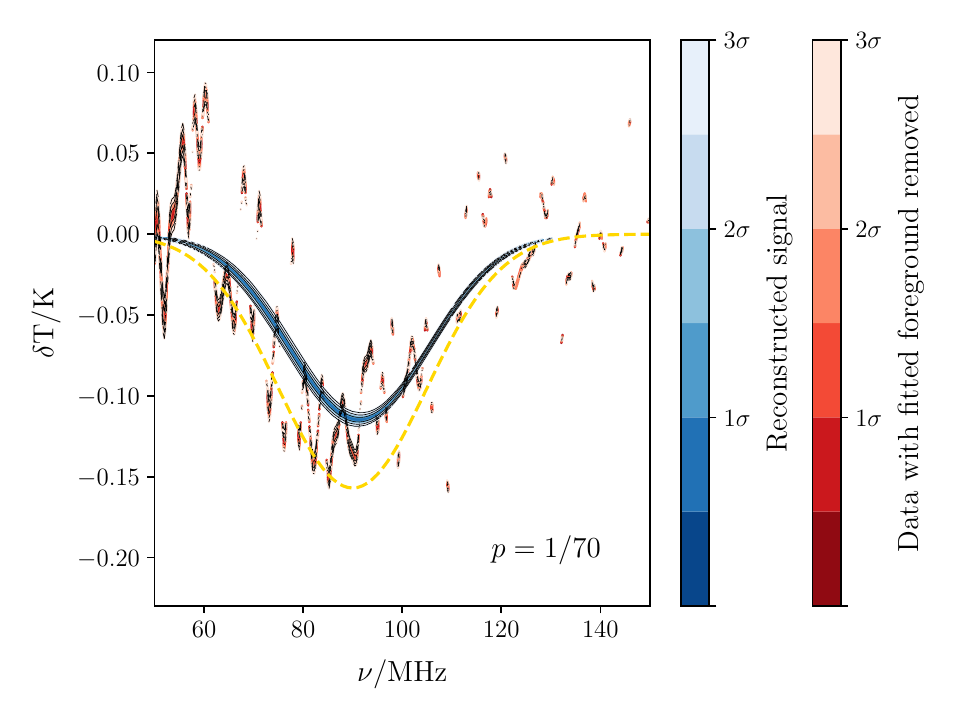}
    \endminipage\hfill
    \caption{Reconstructed signal and foreground residual plots for cases with spatially uniform linear polarisation fraction $p$ of $1/10$, $1/30$, and $1/70$ from left to right respectively. The same polarisation fraction is applied to the whole sky at all frequencies in each case. The Faraday depth map (middle panel in figure \ref{figregs}) used to generate the data is based on the RMTable2023 catalogue \citep{vaneck23}. The reconstructed signal is represented by the blue contours and the foreground residual by red contours. The true signal is marked by the yellow dashed line. Signal reconstruction improves significantly with decreasing polarisation fraction.
    }
    \label{figrmt2023}
\end{figure*}

\begin{figure}
    \centering
    \minipage{0.49\textwidth} \includegraphics[trim={0cm 0cm 0 1cm},clip,width=\linewidth]{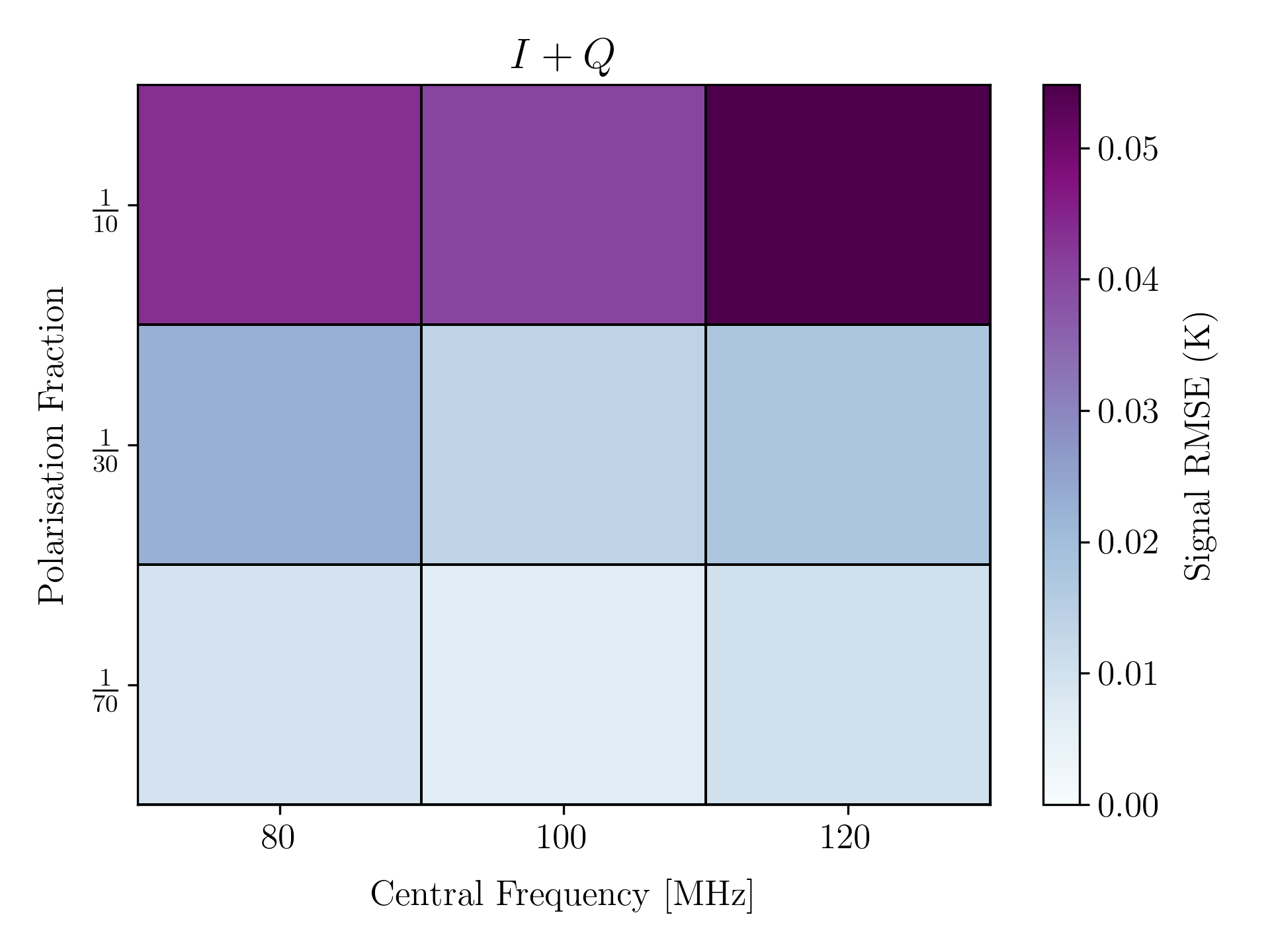}
    \endminipage\hfill
    \caption{The signal RMSE for cases with different polarisation fractions and injected signals of amplitude 0.157 K and width 10 MHz, centred at 80, 100, and 120 MHz respectively. As expected, signal recovery significantly improves with decreasing polarisation fraction across all frequencies. There is not an obvious difference between cases with injected signals centred at different frequencies. The details about this can be found in appendix \ref{append2}.}
    \label{figrmsemeshrot}
\end{figure}

Figure \ref{figrmt2023} shows the reconstructed signal and foreground residual plots for cases with uniform linear polarisation fraction $p$ of $1/10$, $1/30$, and $1/70$, with literature  suggesting that polarisation fraction for the radio sources at 189 MHz have an upper limit of $\sim 2\%$ \citep{bern13}. These values are around the critical level at which one can observe the  quality transition in signal recovery. The injected signal has an amplitude of 0.157 K, a width of 15 MHz and is centred at 90 MHz. For the case with $p = 1/10$, the REACH pipeline completely misses the signal. Whereas the signal is recovered with a diminished amplitude in the other two cases in which the linear polarisation fraction is low. Figure \ref{figrmsemeshrot} shows the signal RMSE for  injected signals of 0.157 K and a width of 15 MHz centred at different frequencies, namely, 80, 100, and 120 MHz. Signal recovery significantly improves with decreasing polarisation fraction across all frequencies. Signal RMSE of less than $10\%$ of the injected signal absorption strength can be achieved for linear polarisation fractions of $1/30$ and $1/10$. The results do not suggest that there is a frequency range more susceptible to polarisation contamination than others. The detailed explanation can be found in appendix \ref{append2}.

\subsection{Number of Faraday Depth Regions} 
\begin{figure}
    \centering
    \minipage{0.49\textwidth} \includegraphics[trim={0cm 0cm 0 0cm},clip,width=\linewidth]{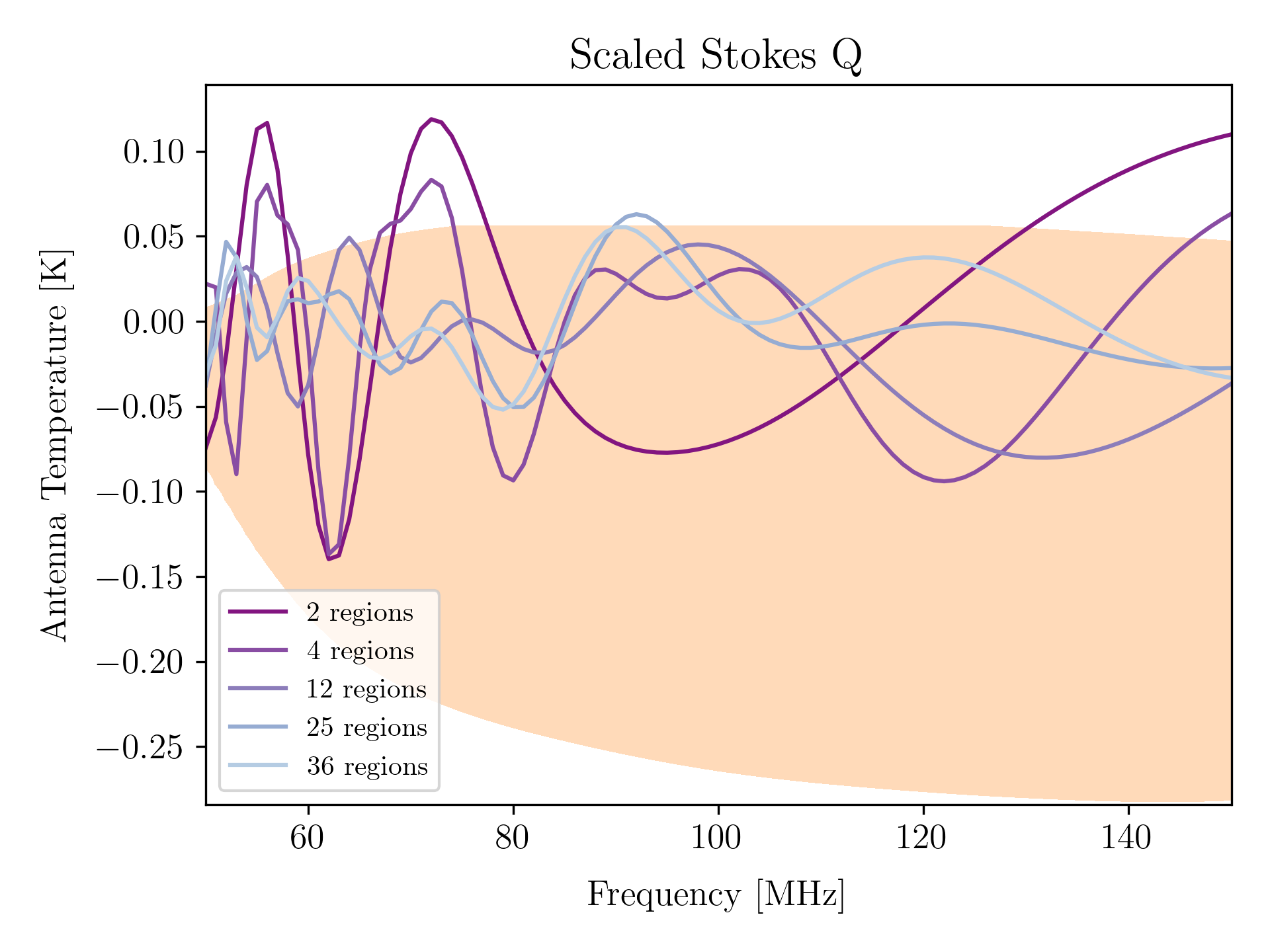}
    \endminipage\hfill
    \caption{Scaled Stokes $Q$ as a function of frequency for different numbers of Faraday depth regions, with a uniform linear polarisation fraction of $p=1/1000$ across all frequencies for all cases. The regions are divided according to the spectral indices using the REACH data analysis pipeline \citep{dom1}. The light-orange shading represents the prior range for a standard global 21-cm signal within 3 sigmas, and the solid lines show the Stokes $Q$ oscillation for cases divided into different numbers of regions. The Stokes $Q$ parameter map is simulated according to equation (\ref{eqqip}). The oscillation of Stokes $Q$ is slower and smoother when the number of regions is low, with its magnitude being comparable to the injected global signal.}
    \label{figqsig}
\end{figure}

\begin{figure*}
    \centering
    \minipage{0.99\textwidth} \includegraphics[trim={0 0 5cm 0cm},clip,height=6.2cm]{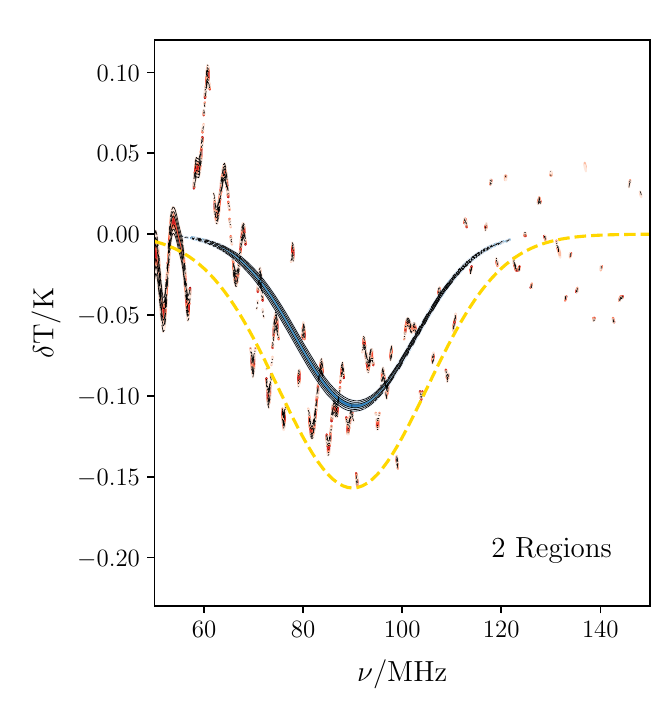}
    \includegraphics[trim={1.3cm 0 5cm 0cm},clip,height=6.2cm]{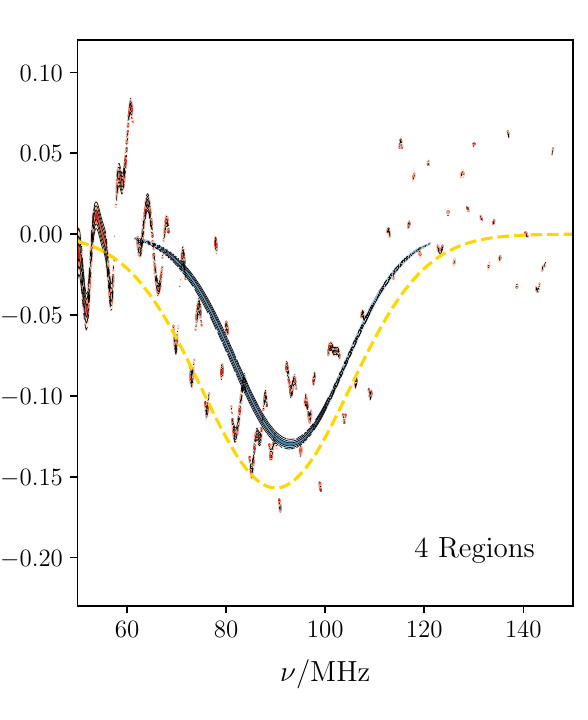}
    \includegraphics[trim={1.3cm 0 0.7cm 0cm},clip,height=6.2cm]{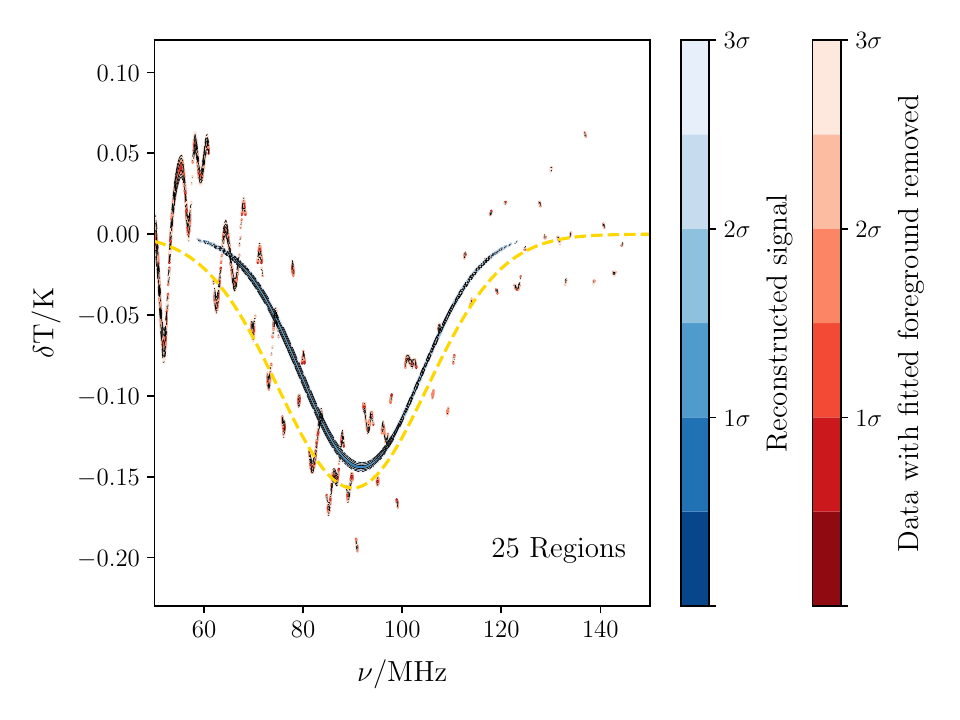}
    \endminipage\hfill
    \caption{Reconstructed signal and foreground residual plots for cases with 2, 4, and 25 Faraday regions in which Faraday depth values are randomly assigned between 0 and 0.5 rad m$^{-2}$. The regions are divided based on the spectral indices across the sky. The homogeneous linear polarisation fraction is $p=1/3000$ for all frequencies. The reconstructed signal is represented by the blue contours and the foreground residual by red contours. The true signal is marked by the yellow dashed line. Signal recovery improves with the increasing number of regions, with progressively lower signal RMSE's and lower residuals.}
    \label{figrespl}
\end{figure*}

Initially, we examined cases in which the regions are divided based on their similarity in spectral indices (shown on the right-hand side of figure \ref{figregs}). It is the sky division scheme adopted in the REACH data analysis pipeline for foreground modelling. In figure \ref{figqsig}, Stokes $Q$ is shown as a function of frequency with a uniform linear polarisation fraction of $p=1/1000$  across the whole sky at all frequencies. The value of polarisation is chosen so that it yields a structure comparable to a standard global signal. Each curve corresponds to a case with a distinct number of Faraday depth regions obtained by spatially integrating over the entire Stokes $Q$ parameter maps simulated at each frequency according to equation (\ref{eqqip}). The oscillation of Stokes $Q$ as a function of frequency (figure \ref{figrespl}) is slower and smoother when the number of regions is low, with its overall magnitude being comparable to the injected global signal.
 
\begin{figure}
    \centering
    \minipage{0.49\textwidth} \includegraphics[trim={0cm 0cm 0 0cm},clip,width=\linewidth]{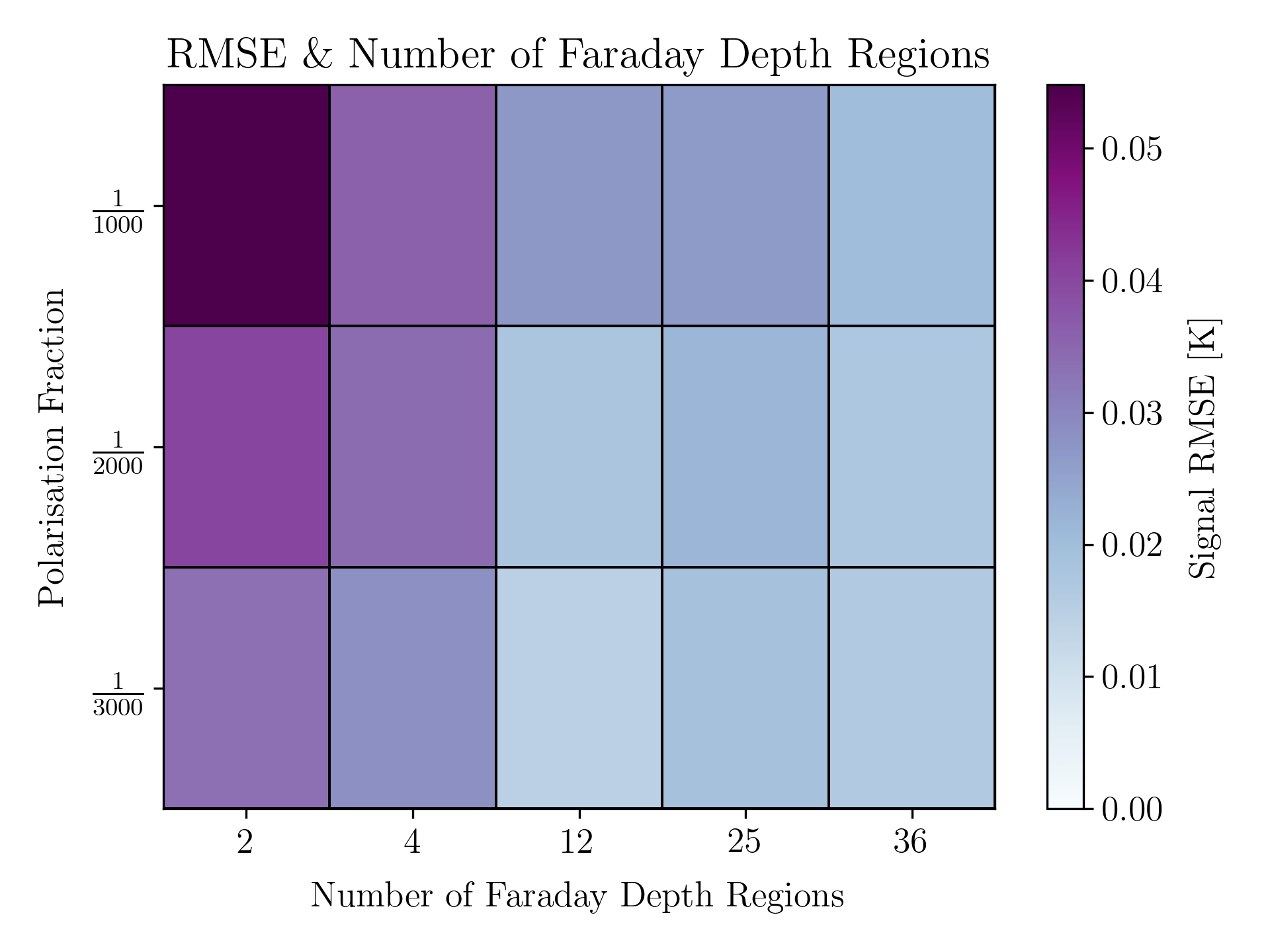}
    \endminipage\hfill
    \caption{The signal RMSE for cases with different numbers of Faraday depth regions for an injected signal of amplitude 0.157 K, width 15 MHz, and centred at 90 MHz. The regions are divided based on the spectral index across the sky, as described in \citet{dom1}. The signal RMSE decreases with the increasing number of regions due to the total polarisation becoming smaller and noisier, making the relatively smoother injected signal easier to distinguish and progressively reducing the RMSE.}
    \label{figrmsemesh}
\end{figure}

 Figure \ref{figrmsemesh} shows the signal RMSE of cases with different numbers of regions and different polarisation fractions. A lower signal RMSE indicates better signal recovery. One can observe that signal recovery improves with decreasing polarisation fraction and increasing number of Faraday depth regions; the former is self-explanatory, as the magnitude of the contamination has a linear dependency on the polarisation fraction (equation \ref{eqqip}). The negative correlation with the number of Faraday regions is due to the  lowering magnitude and faster oscillatory behaviour yielded by the increasing number of regions. Oscillatory behaviour has a greater impact on signal recovery than magnitude; the contamination given by the case of 36 Faraday regions with  $p=1/1000$ is of a similar magnitude as that of 2 regions with $p=1/3000$, and the signal RMSE is significantly lower in the former case.

\section{Conclusions}
\label{seccon}
Contamination from a polarised sky poses a potential challenge to global 21-cm experiments. In this work, three different approaches are adopted to simulate foregrounds with linear polarisation leakage observed by a dipole beam. Simulated effective antenna temperature data, including an injected global signal, are analysed using REACH's time-separated data analysis pipeline, which jointly fits data across all time bins. REACH's data analysis pipeline is built for the purpose of reconstructing the redshifted global signal based on physically motivated foreground models, and the time-separated analysis is designed to improve accuracy by incorporating additional information about the foregrounds and chromaticity.

The first approach employs rotation measure synthesis technique to generate an all-sky Stokes $Q$ sky map for each frequency in the observed range (50 to 150 MHz with 1 MHz resolution) along the line of sight. The model is based on the extrapolation from the MWA 2400 degree$^2$ survey at 189 MHz and simulates the polarised Galactic diffuse emission. The resulting contamination structure agrees with those reported by \citet{spin2018} and \citet{spin2019}, that `low $\phi$' contamination oscillates at a rate that makes it resemble the global signal better, and therefore makes it more challenging to separate. On the other hand, `all $\phi$' cases exhibit fast oscillatory structures that are easier to separate. Nevertheless, the time-separated REACH data analysis pipeline successfully distinguishes the signal in both `low $\phi$' and `all $\phi$' cases, achieving low signal RMSE and high Bayes factors. In \citet{spin2018} and \citet{spin2019}, the standard pipeline struggled under the same amount of contamination; they had to reduce the amplitude of the contamination to get similarly decent results.

The second approach uses the rotation measure data of discrete point sources from the RMTable2023 catalogue \citep{vaneck23}to generate an interpolated all-sky Faraday depth map using nearest-neighbour interpolation." Three Stokes $Q$ sky maps,  each with a different linear polarisation fraction applied uniformly across the sky at all frequencies, are generated to model foreground maps with varying levels of polarisation contamination. The resulting Stokes $Q$ maps have fast oscillatory structures. For injected signals centred between 80 and 120 MHz, signal recovery by the time-separated REACH pipeline improves with decreasing polarisation fraction.  A signal RMSE of less than $10\%$ of the injected signal strength is achievable for linear polarisation fractions of $\leq 3\%$. \citet{bern13} found that polarisation fraction for the radio sources at 189 MHz in their catalogue has an upper limit of $\sim 2\%$. This suggests that the time-separated REACH data analysis pipeline is capable of separating the global signal in the presence of polarisation contamination due to discrete radio sources.

The third approach uses a toy model that divides the sky into regions with randomly assigned low Faraday depth values to test the impact of complex Faraday depth distributions. Linear mixing of contamination oscillating at different rates across frequencies produces unsmooth, relatively noisy structures compared to the foreground. The results indicate that increased variation and spatial complexity in Faraday depth values produce structures more distinct from the global signal, making contamination easier to separate. 

Overall, by testing the three toy models, we find that the time-separated REACH data analysis pipeline is able to recover injected global 21-cm signals of an amplitude of 0.157 K centred between 80 and 120 MHz in the presence of unaccounted-for linear polarisation contamination in the majority of cases. The REACH pipeline significantly outperforms standard approaches, which typically lack the robustness against such contamination. Even in scenarios of unexpectedly high polarisation fractions or complex Faraday depth structures, our results establish clear bounds on the pipeline’s performance.

\section*{Acknowledgements}
ES is supported by Cambridge Trust and Taiwan Ministry of Education. DA is supported by STFC and EdLA is supported by STFC Ernest Rutherford Fellowship. 

\section*{Data Availability}
The data underlying this article will be shared on reasonable request to the corresponding author.



\bibliographystyle{mnras}
\bibliography{ref} 




\appendix

\section{Polarised Foregrounds in Stokes Parameters}
\label{append1}
The total brightness temperature with polarisation leakage observed by an antenna can be described using Stokes parameters, which are directly measurable quantities. In this section, we write down equations relating the polarised foregrounds observed by a dipole antenna to Stokes parameters. More details can also be found in \citet{spin2019}.

\subsection{Electric Field Mean Intensity}
For an electric field described by a pair of perpendicular linear components $(E_x,E_y)$, Stokes $I$ (total intensity including both polarised and unpolarised intensities) and Stokes $Q$ are respectively defined as
\begin{equation}
\begin{aligned}
I &\equiv \langle E_x^2\rangle + \langle E_y^2\rangle, \:\:
Q \equiv \langle E_x^2\rangle - \langle E_y^2\rangle.
\end{aligned}
\label{eqstokesiq}
\end{equation}
The mean intensities of the electric field components $(E_x,E_y)$ can then be derived:
\begin{equation}
\begin{aligned}
\langle E_x^2\rangle &= \frac{I+Q}{2}, \:\: \langle E_y^2\rangle = \frac{I-Q}{2}.
\end{aligned}
\label{eqstokesiqi}
\end{equation}
Thus, given both Stokes $I$ and $Q$ maps, the observed linearly polarised foreground temperature can be simulated using equation (\ref{eqstokesiqi}). Stokes $I$ is the intrinsic intensity, and Stokes $Q$ is the leakage due to polarisation. Since the REACH dipole antenna has only one receptor, one of the electric field components is effectively null in practice.

\subsection{Jones Matrix Representation}
The intrinsic brightness temperature $\mathbf{s}$ along a line of sight $\mathbf{\hat{r}}$,  expressed in terms of Stokes parameters $I$, $Q$, $U$, and $V$, is: $\mathbf{s} = (I,Q,U,V)^\mathrm{T}$, and the brightness temperature observed by the receptor $\mathbf{e} = (\mathcal{E}_{xx},\mathcal{E}_{yy})^\mathrm{T}$ can be written as
\begin{equation}
\begin{aligned}
\mathbf{e}(\mathbf{\hat{r}},\nu)=[\mathbb{J}(\mathbf{\hat{r}},\nu)\otimes \mathbb{J}^*(\mathbf{\hat{r}},\nu)]\mathbb{S}\:\mathbf{s}(\mathbf{\hat{r}},\nu),
\end{aligned}
\label{eqorth}
\end{equation}
where $\mathbb{J}\otimes \mathbb{J}^*$ is the outer product of the $ 2 \times 2$ Jones matrix representing the polarised receptor response and its complex conjugate, while $\mathbb{S}$ is the matrix that transforms the Stokes parameters to the antenna feed frame:
\begin{equation}
\begin{aligned}
\mathbb{S} = \frac{1}{2}\begin{pmatrix}
1 & 1 & 0 & 0\\
1 & -1 & 0 & 0
\end{pmatrix}.
\end{aligned}
\label{eqsss}
\end{equation}
A linearly polarised antenna can be described by a Jones matrix:
\begin{equation}
\begin{aligned}
\mathbb{J} = J\begin{pmatrix}
1 & 0\\
0   & 0
\end{pmatrix},
\end{aligned}
\label{eqjml}
\end{equation}
and by equation (\ref{eqorth}), one gets
\begin{equation}
\begin{aligned}
\mathcal{E}_{xx}(\mathbf{\hat{r}},\nu) &= \frac{1}{2} J_x^2(\mathbf{\hat{r}},\nu)[I(\mathbf{\hat{r}},\nu,t)+Q(\mathbf{\hat{r}},\nu,t)].
\end{aligned}
\label{eqspolexy}
\end{equation}
By defining $A_x = J_x^2/2$, $T_{xx}$, the brightness temperature measured by the receptor $\mathcal{E}_{xx}$ is expressed as:
\begin{equation}
\begin{aligned}
T_{xx}(\mathbf{\hat{r}_0},\nu,t) &\equiv \frac{\int_\Omega \mathcal{E}_{xx}(\mathbf{\hat{r}}',\nu)d\mathbf{\hat{r}}'}{\int_\Omega A_{x}(\mathbf{\hat{r}}',\nu)d\mathbf{\hat{r}}'} \\
&= T_\mathrm{f}(\mathbf{\hat{r}_0},\nu,t) + T_Q(\mathbf{\hat{r}_0},\nu,t) + T_{\mathrm{21}}(\nu),
\end{aligned}
\label{eqstxxtyy}
\end{equation}
where  $T_\mathrm{f}$ is the total foreground intensity, $T_Q$ is the linear foreground polarisation contribution by Stokes $Q$, and $T_\mathrm{21}$ is the true global 21-cm signal.

\section{Difference between Two Orthogonally Pointing Receivers}
\label{append2}
\begin{figure*}
        \minipage{0.99\textwidth} \includegraphics[trim={0cm 0 3cm 0cm},clip,height=7.5cm]{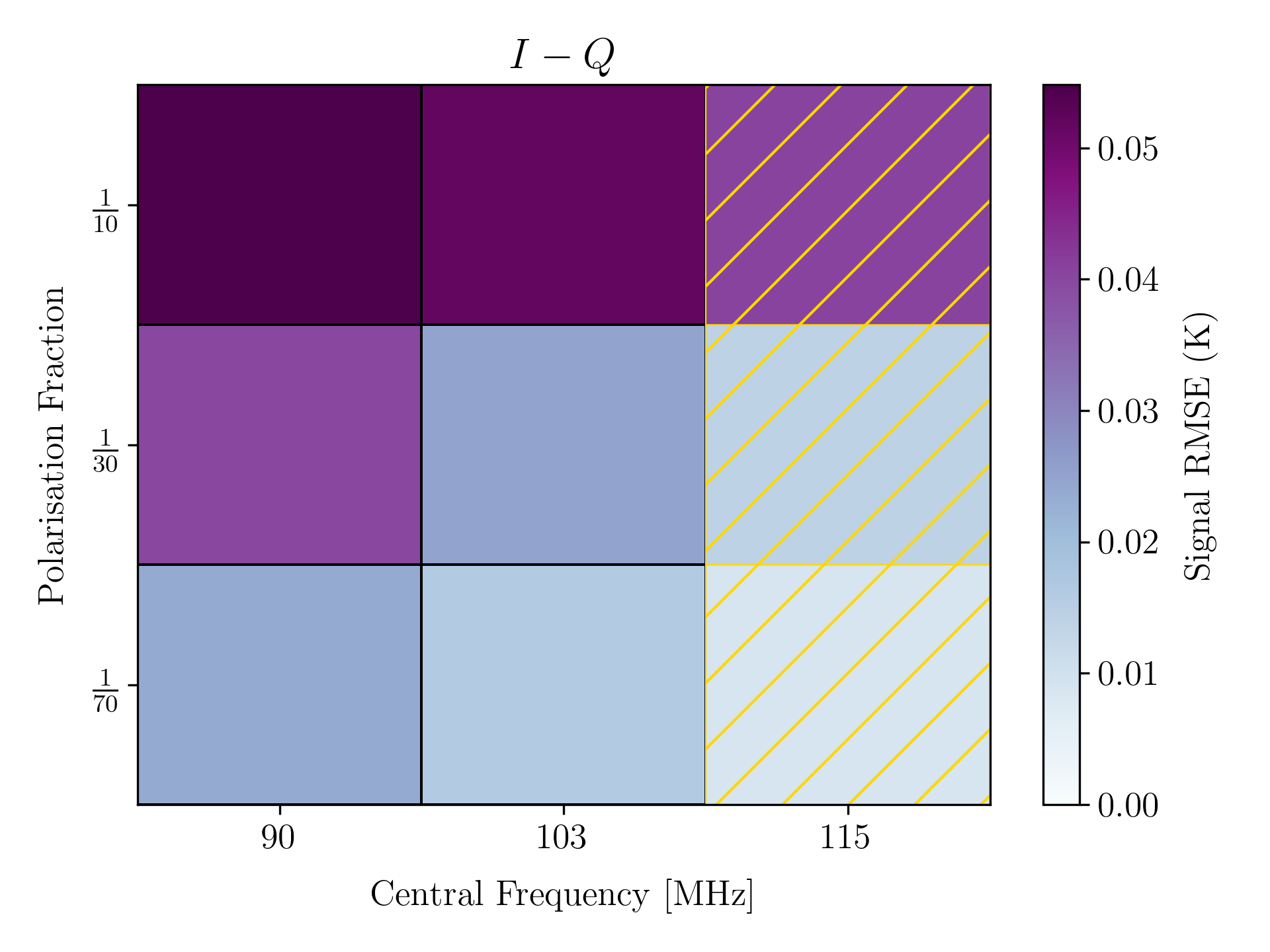}
        \includegraphics[trim={1.2cm 0 0cm 0cm},clip,height=7.5cm]{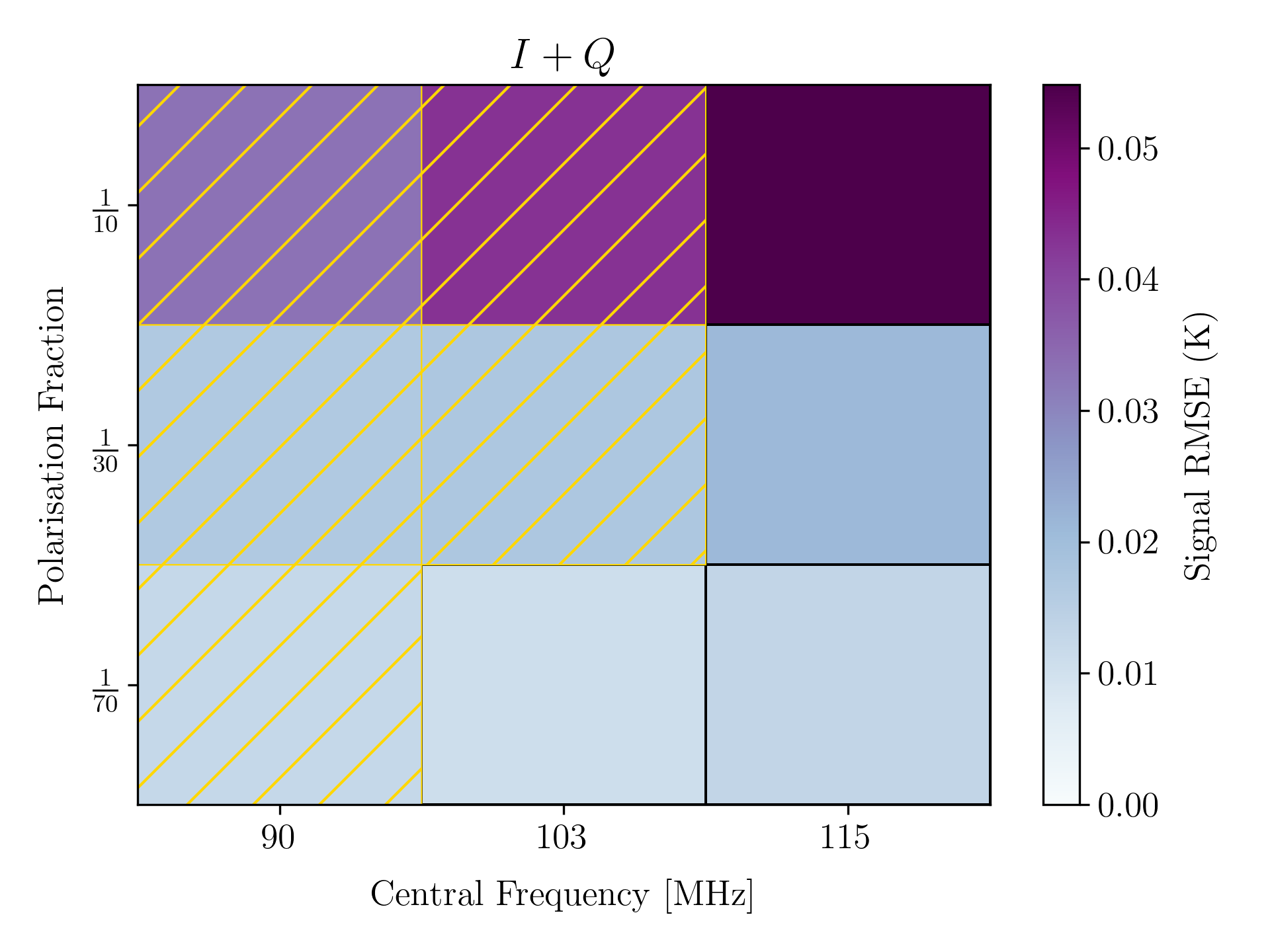}
        \endminipage\hfill\\
        \centering
        \minipage{0.99\textwidth} \includegraphics[trim={0cm 0 3cm 0cm},clip,height=7.5cm]{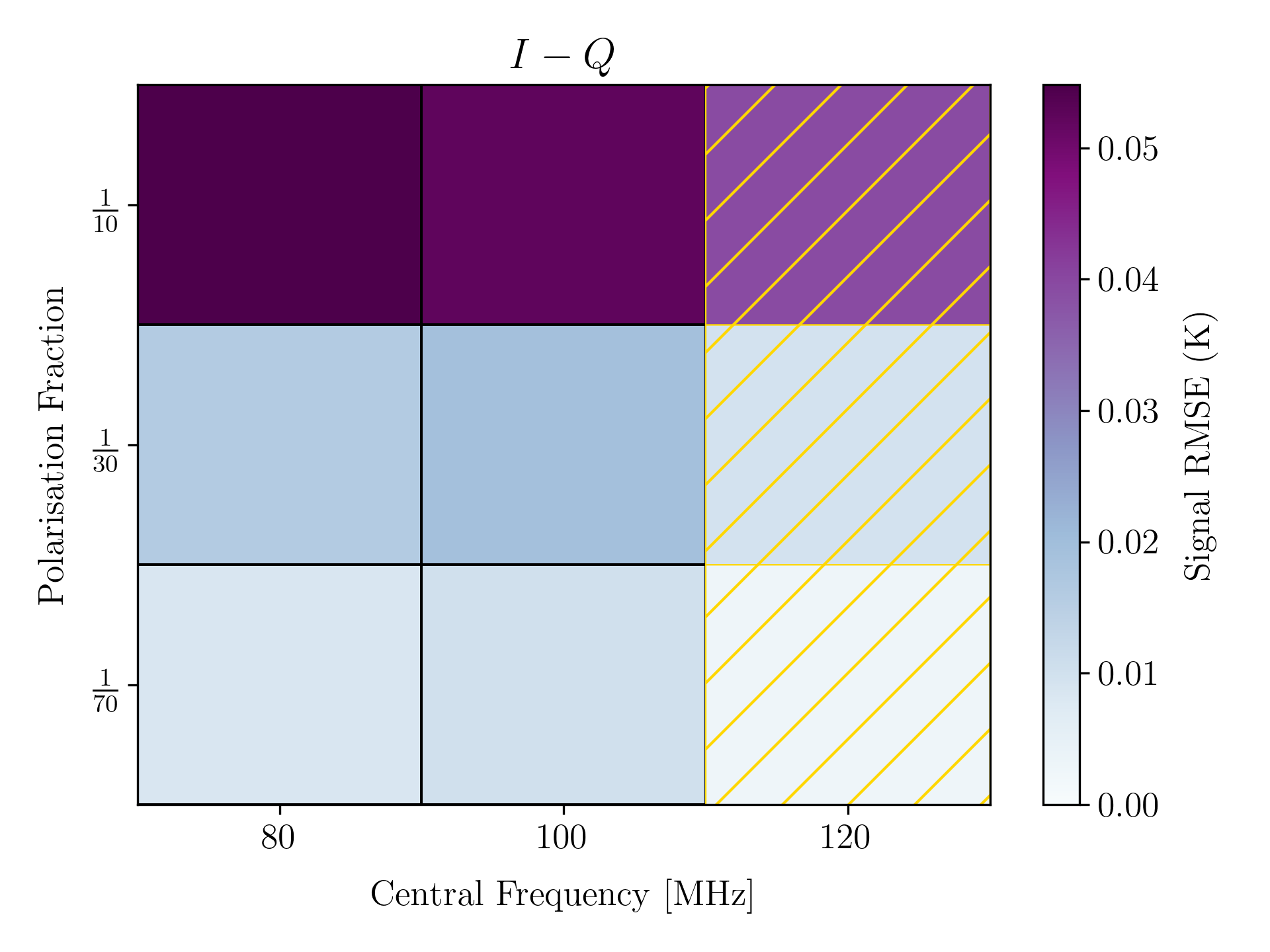}
        \includegraphics[trim={1.2cm 0 0cm 0cm},clip,height=7.5cm]{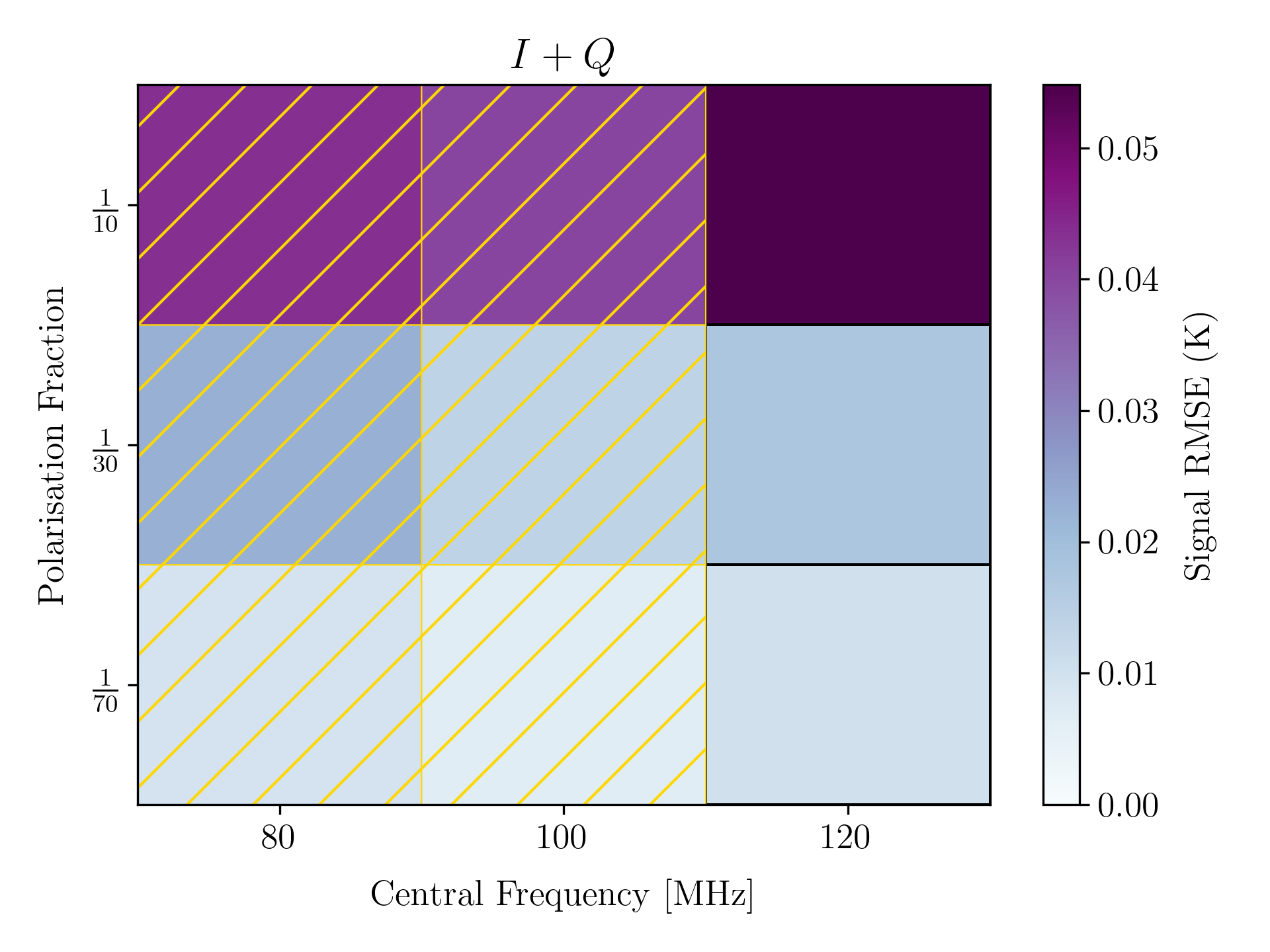}
        \endminipage\hfill
    \caption{The signal RMSE for cases with an injected signal of (top row) amplitude 0.157 K and width 15 MHz centred at different frequencies: 90, 103, and 115 MHz (bottom row) amplitude 0.157 K and width 10 MHz centred at different frequencies: 80, 100, and 120 MHz. Polarisation is simulated by $I+Q$ for the left panel and $I-Q$ for the right panel, corresponding to two orthogonal receivers. Yellow hatches indicate an overestimation of amplitude in the recovered signal. As anticipated, signal recovery improves significantly as the polarisation fraction decreases. No significant difference in signal recovery is observed between injected signals centred at different frequencies or those with differing widths.}
    \label{figrotmesh2}
\end{figure*}

Figure \ref{figrotmesh2} shows the signal RMSE for cases with injected signal centred at different frequencies using two receivers whose directions of pointing are orthogonal to each other. The initial purpose is to determine whether certain frequency ranges are more susceptible to fast-oscillating contamination. Both $I-Q$ and $I+Q$  cases are analysed to observe comparable effects in both directions. $I-Q$ and $I+Q$ are equivalent to two receivers orthogonal to each other. The REACH data analysis pipeline is known to underestimate, rather than overestimate, the depth of the 21-cm signal when using the dipole beam in cases where foregrounds were inadequately modelled. This mitigates signal overestimation in cases where overall contamination enhances the absorption feature of the global signal, as reflected in our results: the signal RMSE is relatively lower than its underestimated counterpart when it is overestimated that the relative degree of accuracy in signal recovery is almost the opposite between the $I-Q$ and $I+Q$ cases. This behaviour suggests that underestimation or overestimation is the primary driver of the relative signal RMSE difference, rather than any specific frequency range being particularly susceptible to polarisation contamination.


\bsp	
\label{lastpage}
\end{document}